\documentclass[preprint,emulateapj,twocolumn]{aastex}
\usepackage{epsf}
\usepackage{graphicx,color}
\usepackage{emulateapj5}
\usepackage{onecolfloat}
\usepackage{apjfonts}
\usepackage{amsmath}

\newcommand{\lsim}{\lower0.6ex\vbox{\hbox{$ \buildrel{\textstyle <}\over{\sim}\ $}}}
\newcommand{\gsim}{\lower0.6ex\vbox{\hbox{$ \buildrel{\textstyle >}\over{\sim}\ $}}}

\newcommand{\black}{\color{black}}
\newcommand{\white}{\color{white}}

\newcommand{\pink}{\color{\pink}}

\newcommand{\msun}{M_{\odot}}

\newcommand{\pc}{{\,{\rm pc}}}
\newcommand{\kpc}{{\,{\rm kpc}}}

\newcommand{\half}{\frac{1}{2}}

\newcommand{\Mvir}{M_{\mathrm{vir}}}

\newcommand{\beq}{\begin{equation}}
\newcommand{\eeq}{\end{equation}}

\newcommand{\Gammai}{\Gamma_{\mathrm{i}}}

\newcommand{\avg}[1]{\langle #1 \rangle}
\newcommand{\Var}[1]{\mathrm{VAR}( #1 )}

\newcommand{\bmath}[1]{\mbox{\boldmath ${#1}$}} 
\newcommand{\btheta}{\bmath{\theta}}
\newcommand{\bthetap}{\bmath{\theta}_{\mathrm{p}}}
\newcommand{\dbtheta}{\delta \bmath{\theta}}

\newcommand{\bbeta}{\bmath{\beta}}
\newcommand{\balpha}{\bmath{\alpha}}
\newcommand{\dbalpha}{\delta \bmath{\alpha}}

\newcommand{\eunit}{{\bf e}}

\newcommand{\zl}{z_{\mathrm{l}}}
\newcommand{\zs}{z_{\mathrm{s}}}
\newcommand{\dl}{D_{\mathrm{L}}}
\newcommand{\ds}{D_{\mathrm{S}}}
\newcommand{\dls}{D_{\mathrm{LS}}}

\newcommand{\Sc}{\Sigma_{\mathrm{c}}}

\newcommand{\Sigmas}{\Sigma_{\mathrm{s}}}

\newcommand{\tb}{\tilde{b}}

\newcommand{\fsat}{f_{\mathrm{sub}}}
\newcommand{\dt}{\delta_{\mathrm{T}}}
\newcommand{\drms}{\delta_{\mathrm{T}}^{\mathrm{rms}}}
\newcommand{\dirac}{\delta_{\mathrm{D}}}

\newcommand{\tmin}{\theta_{\mathrm{min}}}
\newcommand{\tmax}{\theta_{\mathrm{max}}}
\newcommand{\phim}{\phi_{\mathrm{m}}}

\newcommand{\dnl}{\Delta_{\mathrm{NL}}}
\newcommand{\dmin}{\delta_{\mathrm{min}}}
\newcommand{\deltap}{\delta_{\mathrm{p}}}

\newcommand{\mmax}{m_{\mathrm{max}}}
\newcommand{\mmin}{m_{\mathrm{min}}}

\newcommand{\fnl}{f_{\mathrm{NL}}}
\newcommand{\fl}{f_{\mathrm{L}}}

\newcommand{\mand}{\mbox{ and }}

\newcommand{\mspaces}{\hspace{0.5 in}}

\begin{document}


\submitted{The Astrophysical Journal, submitted}
\vspace{1mm}
\slugcomment{{\em The Astrophysical Journal, submitted}}

\shortauthors{Rozo et al.}


\twocolumn[
\lefthead{Magnification Perturbations in CDM Cosmologies}
\righthead{Rozo, Zentner, Bertone, \& Chen}

\title{Statistics of Magnification Perturbations 
by Substructure in the Cold Dark Matter Cosmological Model}

\author{
Eduardo Rozo\altaffilmark{1,2},
Andrew R. Zentner\altaffilmark{2,3},
Gianfranco Bertone\altaffilmark{4},
and Jacqueline Chen\altaffilmark{2,3}
}

\begin{abstract}

We study the statistical properties of magnification perturbations by
substructures in strong lensed systems using linear perturbation theory 
and an analytical substructure model including
tidal truncation and a continuous substructure mass spectrum.
We demonstrate that magnification perturbations are dominated
by perturbers found within roughly a tidal radius of an image, and 
that sizable magnification perturbations may arise from small, 
coherent contributions from several substructures within the 
lens halo.  The root-mean-square (rms) fluctuation of the 
magnification perturbation is $\sim 10\%$ to $\sim 20\%$ 
and both the average and rms perturbations are sensitive to the 
mass spectrum and density profile of the perturbers.
Interestingly, we find that relative to a smooth model of the same
mass, the average magnification in clumpy models is 
{\em lower} ({\em higher}) than that in smooth models 
for positive (negative) parity images.
This is opposite from what is observed if one assumes that the
image magnification predicted by the best-fit smooth model of a 
lens is a good proxy for what the observed magnification would 
have been if substructures were absent.  
While it is possible for this discrepancy to be 
resolved via nonlinear perturbers, we argue that a more likely 
explanation is that the assumption that the best-fit lens model 
is a good proxy for the magnification in the absence of 
substructure is not correct.  We conclude that a better theoretical
understanding of the predicted statistical properties of magnification 
perturbations by CDM substructure is needed in order to affirm that 
CDM substructures have been unambiguously detected. 

\end{abstract}

\keywords{cosmology: theory -- dark matter -- 
galaxies: formation, halos, structure}
]


\altaffiltext{1}{
Department of Physics, The University of Chicago, 
Chicago, IL 60637, USA
{\tt erozo@oddjob.uchicago.edu}
}

\altaffiltext{2}{
Kavli Institute for Cosmological Physics, 
Chicago, IL 60637, USA
}

\altaffiltext{3}{
Department of Astronomy and Astrophysics, 
The University of Chicago, 
Chicago, IL 60637, USA
}

\altaffiltext{4}{
NASA/Fermilab Astrophysics Center,
Fermi National Accelerator Laboratory, 
Batavia, IL 60510, USA
}


%
%

\section{Introduction}
\label{sec:intro}

\subsection{Evidence for Small-Scale Features in Strong Lenses}

Standard lens models often have difficulties explaining 
the relative fluxes of multiply-imaged sources 
\citep[][]{MetcalfZhao02}.  
Perhaps the simplest explanation for these discrepancies is 
that they are electromagnetic in origin.  
Observed fluxes may be affected by obscuration due to 
dust, scintillation in the galaxy, or some other form of 
electromagnetic phenomena 
\citep[e.g.,][]{Koopmansetal03}.  
However, there is strong evidence that these are not 
the only source of the discrepancies.  
In particular, \citet{KochanekDalal04} 
have shown that positive-parity images in observed lenses
tend to be brighter than what the standard lens models predict, 
while negative-parity images tend to be demagnified.  
Electromagnetic effects do not distinguish 
between positive-parity and negative-parity images, 
so we are left to conclude that flux anomalies must 
be produced by gravity.  That is,
the lensing potentials of real galaxies 
are not fully characterized by the simple 
models used to describe them.

The next simplest explanation for these discrepancies 
is that standard lens models are either overly simplistic 
or overly restrictive 
\citep[]
[though see also \citeauthor{Yooetal05} \citeyear{Yooetal05}]
{Kawanoetal04,EvansWitt03}.  
For instance, environmental effects are often
modeled as a constant, external shear, 
whereas detailed modeling
of the lens environment might be necessary 
\citep{Molleretal02, KeetonZabludoff04}.  
Again, there is strong evidence that this is not 
the only difficulty.

The first piece of evidence that other difficulties must 
exist concerns the so-called {\em cusp relation}.
Cusp lens configurations occur when a source is located 
near one of the cusps of a lens' tangential caustic. They 
characteristically have a tight cluster of three 
images and a significantly more isolated fourth 
image.\footnote{A fourth image may be absent in the 
so-called ``naked cusp" lenses, which occur when the 
cusp of a tangential caustic is not contained in the 
region interior to the radial caustic in the plane of the source.}  
The cusp relation relates the signed fluxes $\Gammai$, 
of each of the images (labeled by index $\mathrm{i}$), where 
the sign of the flux is given by the parity of the image.  
For cusp configurations, one can define the cusp parameter 
$R = \sum_{\mathrm{i}} \Gammai/(\sum_{\mathrm{i}} |\Gammai|)$, 
where the sum is over the three tightly-clustered images.
For a point source located near the cusp, one expects to find 
$R \approx 0$ so long as the lensing potential is 
featureless on scales comparable to or smaller than 
the typical image separations 
\citep{SchneiderWeiss92, Zakharov95, GaudiPetters02b}. 
\citet{KeetonGaudiPetters} have shown that 
careful analysis can identify lenses where the condition 
$R\approx 0$ is violated, providing strong evidence that 
small-scale features in the lensing potential are present.  
A similar argument can be made for fold lenses where 
one expects a close pair of images with equal and 
opposite fluxes in addition to two isolated images.  
This fold relation is likewise violated in a large 
number of lenses 
\citep[][]{KeetonGaudiPetters05b}.  
Note that while such detections are  
model independent and robust,
they require further model-dependent 
analysis to determine possible causes of 
the perturbation and to 
identify which image (or images) is (are) perturbed 
\citep[e.g.,][]{MaoSchneider98, Keeton01, DoblerKeeton05}.

A second piece of evidence for small-scale structure in 
lenses is the frequency dependence of the flux ratios of 
multiply-imaged sources \citep{MoustakasMetcalf03}.  
This argument hinges on the fact that features in the 
lensing potential smaller than the source size tend to be
smoothed out by the source. 
As an illustration, consider two spatially-coincident 
sources of known brightness, one large and one small, 
lensed by an intervening lens 
potential with features on scales intermediate between the 
two source sizes. In such a case, 
the smaller source is strongly affected by 
these intermediate features, 
whereas the larger source is not because the observed image 
flux constitutes an average over a region larger than the 
scale of the features.  Therefore, the existence of the 
features is signaled through different observed magnifications 
for each of the two sources.  Fortunately, 
Nature has been kind enough to 
provide exactly this kind of setup.  
In particular, the spatial extent 
of the emission region of a quasar 
depends on the particular
frequency at which the quasar is observed 
\citep[e.g.,][]{KembhaviNarlikar99}.  By comparing flux ratios of 
multiply-imaged quasars in different spectral regions, one 
may search for structure in the lens potential on length scales 
intermediate between those of the quasar emission regions.  
Again, more detailed modeling is necessary 
to determine the nature of the perturbation
\citep{Metcalfetal04}.

\subsection{CDM Substructure as a Possible Source for Small-Scale
Features in the Lensing Potential}

To our knowledge, \citet{MaoSchneider98} were the first  
to propose that intermediate-mass-scale substructures within 
a lens could explain the problem posed by anomalous flux ratios.  
The Cold Dark Matter (CDM) paradigm of cosmological 
structure formation \citep[e.g.,][]{blumenthal_etal84, white_rees78} 
predicts just such substructure.  In recent years, a new 
generation of numerical simulations of structure 
formation in the CDM paradigm revealed that the 
dark matter halos that are believed to 
host galaxies generally have $\sim 10\%$ of their 
mass in distinct, gravitationally-bound, 
substructures commonly referred to as 
{\em subhalos} or {\em satellite halos} 
\citep[e.g.,][]{Klypinetal99, Mooreetal99, 
ghigna_etal00, delucia_etal04, 
diemand_etal04, gao_etal04a}, 
and it was quickly realized that these subhalos are 
an ideal candidate for the source of 
the lensing perturbations \citep[see][]{Mooreetal99}.
Conversely, estimates of the abundance of substructure 
can serve as a test of models of cosmological 
structure formation 
\citep[e.g.,][]{DalalKochanek02b,zentner_bullock03}.

The amount of substructure predicted by the CDM model has 
been the subject of numerous recent studies.  
One apparent discrepancy between theory and observation is 
that there are more than an order of magnitude fewer dwarf 
satellite galaxies about the Milky Way and M31 
than the number of subhalos of comparable velocity 
dispersion predicted by the CDM paradigm 
\citep{kauffmann_etal93, Klypinetal99, Mooreetal99}.  
This mismatch is known as the ``missing satellites problem,'' 
and many possible resolutions have been considered.  
Several authors have proposed modifications to the properties 
of the dark matter, making it ``warm'' rather than cold 
\citep{hogan_dalcanton00,colin_etal00, 
bode_etal01,lin_etal01,knebe_etal02} 
or introducing a self interaction for the dark matter 
\citep{spergel_steinhardt00} 
in order to reduce the amount of substructure.  

Alternatively, any suppression in 
the amount of small-scale power in the primordial spectrum of 
density fluctuations can significantly influence 
halo structure \citep{zentner_bullock02,mcgaugh_etal03,vdb_etal03} 
and may reduce the abundance of subhalos 
\citep[e.g.,][]{kamionkowski_liddle00,sigurdson_kamionkowski04}.  
\citet{zentner_bullock03} studied in detail 
the dependence of halo substructure 
on the primordial power spectrum and showed 
that even mild modifications to the power spectrum 
on small scales can greatly affect the 
severity and interpretation of the 
missing satellites problem.  These authors also estimated 
projected subhalo mass fractions for several different 
models, demonstrating how measurements of this quantity 
through lensing flux anomalies 
\citep[e.g.,][]{DalalKochanek02a} 
could provide information about the nature of the 
dark matter and the primordial power spectrum, and 
thus inflation, on scales that may 
be {\em otherwise inaccessible}.

The most conservative solution to the missing 
satellites problem is that star formation 
may be naturally suppressed in a large fraction of 
small halos due to feedback from supernovae
\citep[e.g.,][]{dekel_silk86, kauffmann_etal93, 
cole_etal94, somerville_primack99} 
or from the ionizing background 
\citep[e.g.,][]{rees82, efstathiou92, kauffmann_etal93, 
shapiro_etal94, thoul_weinberg96, 
bullock_etal00, somerville02, benson_etal02}. 
This leads to two possible scenarios.  The 
feedback mechanism could cause an abrupt drop 
in the efficiency of galaxy formation in small 
halos, in which case the MW satellites should lie 
in the eleven most massive subhalos surrounding the 
MW.  This could be true if the mapping between 
observed stellar velocity dispersions of the satellites and 
the size of the subhalos that host these satellites 
allows for the satellites to sit in large subhalos 
due to the significant tidal evolution of the subhalos 
\citep[e.g.][though recent work by 
\citeauthor{kazantzidis_etal04b} 
\citeyear{kazantzidis_etal04b} 
does not support this scenario]
{stoehr_etal02,hayashi_etal03}.
Contrarily, feedback could set in gradually so 
that small subhalos become increasingly less likely 
to host luminous satellites.  
Including several other details, this type of 
scenario leads to a model of biased galaxy formation 
in small halos that appears to agree well with many 
observed features of the MW dwarfs 
\citep{kravtsov_etal04,zentner_etal05a}.

If inefficient galaxy formation in small halos is the 
resolution to the missing satellites problem, 
galaxy-sized subhalos should be filled with massive, 
dark subhalos that would be devoid of stars and 
detectable only through their gravitational influence or, 
more speculatively, through the detection of gamma-rays
\citep[e.g.,][]{silk_stebbins93,berezinsky_etal97,bergstrom_etal99,
calcaneo-roldan_moore00,baltz_etal00,tasitsiomi_olinto02,stoehr_etal03,
koushiappas_etal04} 
or antiprotons \citep[e.g.,][]{bergstrom_etal99,bottino_etal98,donato_etal03}
from dark matter annihilations in their dense cores (for 
further details see \citet{Bertone_etal05} and references therein). 
An unequivocal detection of subhalos through lensing 
could be used to constrain the properties of the dark 
matter and the primordial power spectrum on small scales 
\citep{DalalKochanek02b,zentner_etal05b}.  
A detection at sufficiently high levels could represent 
an enormous triumph for the CDM paradigm of structure 
formation with a standard, scale-invariant primordial 
power spectrum.  Such a detection could help 
determine the nature of the feedback mechanism 
that leads to the dearth of small, 
luminous satellites in the MW and distinguish 
between the proposal of \citet{stoehr_etal02} and 
\citet{hayashi_etal03} and that of 
\citet{kravtsov_etal04}.

It is important to note that, while 
dark matter substructure within the lens halo seems to be 
the leading interpretation of observed flux anomalies, 
it is certainly {\em not} the only possibility.  
Other possible sources of small-scale fluctuations 
in lens potentials include stars, correlated and 
uncorrelated external halos along the line of sight 
\citep{ChenKravtsovKeeton, MollerBlain01, 
Metcalf04a, Metcalf04b, Wambsganssetal04}, 
and possible disk structures within the galaxy 
\citep{Quadrietal03, MollerHewettBlain03, 
Bradacetal04, Amaraetal04}.  
Microlensing by stars in the lens galaxy, 
while expected and observed in some cases 
\cite[e.g.,][]{Schild96, Refsdaletal00, Wozniaketal00, 
Wisotzkietal03, Schechteretal03, Richardsetal04}, 
should not affect the radio or narrow-line emission 
fluxes due to the source size 
\citep{MoustakasMetcalf03,KochanekDalal04,DoblerKeeton05}, 
and cannot be the cause of radio flux anomalies.

In light of this discussion, predicting
the properties of magnification perturbations
due to CDM substructures in realistic lens systems 
is of paramount importance.  
Unfortunately, the detailed CDM predictions for lens 
systems are not yet completely understood.  
The first issue to address is whether or not 
the amount of substructure predicted by 
the standard CDM model is consistent with observed 
flux perturbations.  
This problem has been addressed 
analytically \citep{ChenKravtsovKeeton}, 
numerically \citep{MetcalfMadau01}, 
and observationally 
\citep{DalalKochanek02b, MetcalfZhao02, Chiba02, Bradacetal02} 
with fairly consistent results: 
a projected substructure mass fraction, 
$\fsat \equiv \Sigma_{sub}/\Sigma_{tot}$, 
of order a few percent near the images 
can reproduce the flux perturbations 
in observed gravitational lenses.  
Whether this constitutes agreement with CDM
or not is unclear.  
While the total mass fraction in substructure for CDM
halos is $\sim 10\%$, na{\"{\i}}ve theoretical estimates of 
the projected subhalo mass fraction near typical 
Einstein radii of lenses ($\sim$ a few percent of the 
halo virial radius) are significantly lower, yielding 
$\fsat \lesssim 0.5\%$ 
\citep{zentner_bullock03, Maoetal04}.  
However, it is not clear whether such a na{\"{\i}}ve 
comparison is fair (Chen \& Rozo 2005, in preparation) 
because massive substructures could have important effects
even if they do not fall within the Einstein radius
of the host galaxy.  Further, spatially-biased
substructure distributions will clearly affect the lensing
potential, and to what extent these effects can be subsumed
within the lens model remains unknown.
In related investigations, 
\citet{Bradacetal04} have argued that the predicted 
amount of substructure in lenses is sufficient 
to explain the level of cusp relation violations 
observed in the data using simulations of galaxy formation 
in CDM, though \citet{Amaraetal04} \citep[see also][]{Maccio_etal05} 
reached the opposite conclusion.  In summary, 
strong lensing is an important probe of small-scale 
structure and the question of whether magnification 
perturbations are observed at the level that is 
expected from cold dark matter substructures
has not been answered unambiguously.

\subsection{This Work}

In the present work, we quantify the magnifications 
perturbations predicted by the CDM model of structure 
formation and develop an understanding of the important 
aspects to consider highlighting several key parameters.  
In particular, we predict the statistical properties of 
magnification perturbations due to subhalos in realistic 
lens models.  We begin in \S~\ref{section_cross_sections}
by generalizing the work of \citet{Keeton03}.  In particular,
using linear perturbation theory
we derive a general expression for the cross section of
a single perturber to produce a magnification 
perturbation above some given size assuming only 
circularly-symmetric and monotonically-decreasing 
substructure surface density profiles. 
Appendix~\ref{ptmass} and Appendix~\ref{sis} illustrate our
formalism in the context of point mass and Singular Isothermal
Sphere (SIS) perturbers.  In this context, we highlight the 
importance of tidal truncation on the probability of finding
nonlinear perturbers by showing that the expected number
of nonlinear perturbers increases by an order of magnitude
if the tidal truncation of substructure is ignored.

In \S~\ref{section_multiple_perturbers} we build on these
results to derive expressions for the average and variance 
of the total magnification perturbation due to an 
ensemble of perturbers (not necessarily identical). 
We then apply our results to the specific problem
of magnification perturbations by CDM substructures
within lens galaxies in \S~\ref{substruct}.  
We use a simple, yet realistic, model for halo substructure
which includes tidal truncation and a substructure 
mass spectrum, and thereby predict the
average and variance of the magnification perturbations
generated by CDM substructures.  
This type of argument provides an important 
{\em statistical} test of the CDM paradigm, 
which is particularly relevant given that
the level of substructure near the Einstein 
radius of CDM halos is expected to vary enormously 
from halo to halo 
\citep[e.g.,][]{zentner_bullock03,Maoetal04,
taylor_babul05,zentner_etal05c}.  Furthermore,
our analytical treatment allows us to identify
and isolate the various characteristic of substructures 
to which the statistical properties of magnification 
perturbations are sensitive to, such as substructure
profiles and their mass spectrum.

We find that small, negative values for the average 
magnification perturbation are a generic prediction.  
In other words, positive-parity 
images generated by a lens with substructure
are {\em dimmer} than the corresponding images for
a lens of the same mass with no substructure.  
We show that this can be understood as
a byproduct of requiring perturbations to be linear:  
introduction of a nonlinear cutoff implies that a
small fraction of the mass is ignored, which leads
to a small negative magnification perturbation.
The dimming of positive parity images is opposite to what is observed 
{\em if} one assumes
that the best-fit model for a lens is a good proxy 
for the lensing potential obtained by replacing all
substructures by a smooth component, which we
argue is likely not a good assumption.  In 
\S~\ref{discussion} we summarize our results and 
draw conclusions.  Lastly, in Appendix~\ref{appendixa}, 
we discuss in detail the validity of linear 
perturbation theory, and argue that linear 
perturbation theory should be valid whenever
astrometric perturbations are negligible.
Finally Appendix~\ref{ptmass} and Appendix~\ref{sis}
illustrate our algorithm for computing magnification perturbation
cross sections explicitly in the cases of point mass and singular
isothermal sphere perturbers respectively.
%
%

%
%
\section{Magnification Perturbation Cross Sections}
\label{section_cross_sections}

In this section, we illustrate a general algorithm for computing 
the cross sections $\sigma(\delta)$ for individual perturbers 
to generate  magnification perturbations of size 
$ \delta \equiv \delta \mu / |\mu| $ 
or larger ($\mu$ is the unperturbed image magnification).  
Note that for a constant surface number density of perturbers $s$, 
the probability of finding a perturber that creates a perturbation 
of size $\delta$ is simply $dP = s |d \sigma / d \delta | d \delta $.  
This forms the basis of our analysis of multiple perturbers and 
underlines the importance of understanding the magnification 
perturbation cross section $\sigma ( \delta )$.

Consider the image of a point source lensed by a smooth, projected 
lensing potential $\psi$, which we refer to as the 
{\em macrolens} or {\em macromodel}, 
and let $\kappa$ and $\gamma$ be the 
values of the convergence and shear fields at 
the image position.  
We investigate the magnification perturbation of the image
when one introduces a small perturbation to the potential, $\delta\psi$.  
In general, the perturbation $\delta\psi$ will not only change the 
flux of the image, but also its position.  However, here we 
will work in the limit that astrometric perturbations are 
negligible.  We emphasize that negligible astrometric
perturbations means that the change in the convergence and 
shear values of both the macrolens {\em and the perturber} 
have to be negligible.  As we show in Appendix \ref{appendixa}, 
we expect this assumption to hold for small flux perturbations, 
that is, when
\begin{equation}
\label{eq:dsmall}
|\delta|=\Big|\frac{\delta\mu}{\mu}\Big| \ll 1,
\end{equation}
where the flux perturbation $\delta\mu$ is evaluated at the 
unperturbed image position.  Consequently, we will only be 
considering small perturbations to the flux.  

\begin{figure}[t]
\epsscale{1.0}
\plotone{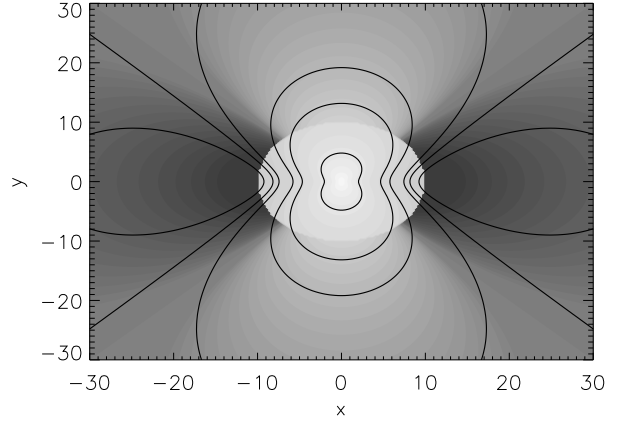}
\caption{
Shown above are the $\delta$ contours around a positive
parity image.  The field $\delta(\bf{x})$ is defined as the
size of the magnification perturbation experienced by an image
at the origin when a perturber
is placed at point $\bf{x}$.  The shaded contours are
obtained assuming truncated SIS perturbers, with dark
being negative and light being positive.  The boundaries 
extending out at $45^\circ$ angles in the shaded contours correspond
to $\delta=0$ (no perturbation).  
Distances are measured in units of the Einstein radius and the
tidal radius is assumed to be ten times the Einstein radius,
clearly seen as a discontinuity
in the shaded contours. The angle $\phi$ is defined relative
to the $x$ axis, itself defined such that the macroshear is
$\bmath{\gamma}=(\gamma,0)$.  The solid contours are obtained
using a Pseudo-Gaffe profile, and correspond to
$\delta=\{-0.01,0.0,0.01,0.05,0.10,0.50\}$.  
The double lobe features of
SIS perturbers found by \citet[][]{Keeton03} are evident 
at distances smaller than the tidal radius.
}
\label{geometry}
\end{figure}

Let $\delta\psi$, $\delta\kappa$, and 
$-\delta \gamma ( \cos(2\phi) , \sin(2\phi) )$ be 
the profiles of the potential, convergence, 
and the two components of the shear of an individual 
perturber.  We assume that all profiles vary monotonically.  
The Poisson equation for the potential $\psi$ is linear,
so the total convergence and shear are given simply by 
$\kappa + \delta \kappa$ and 
$( \gamma - \delta \gamma \cos (2 \phi ), 
-\delta \gamma \sin( 2 \phi ) )$.  
Linearizing the magnification $\mu^{-1} = (1-\kappa)^2-\gamma^2$, 
results in\footnote{The negative sign in front of $\gamma$ in
Equation~(\ref{eq:smallfluxpert}) is due to our sign convention.
One has $\vec{\gamma}=-\gamma(\cos(2\phi),\sin(2\phi))$, so
aligning the shear with the $x$ axis can mean either
$\phi=0$ for $\vec\gamma=(-\gamma,0)$ or $\phi=\pi/2$
for $\vec\gamma=(\gamma,0)$.  We use the latter.}
\begin{equation}
\delta(\theta,\phi) 
= \frac{\delta\mu}{|\mu|} 
= 2|\mu| \left[ (1-\kappa)\delta\kappa(\theta)-
\gamma\delta\gamma(\theta)\cos (2\phi) \right] ,
\label{eq:smallfluxpert}
\end{equation}
where $\kappa$ and $\gamma$ are 
the convergence and shear of the macromodel respectively. 
We have taken the origin to be at the center of 
the perturber, and the coordinates $\theta$ and $\phi$ 
represent two-dimensional polar coordinates.  
In what follows, 
it is more convenient to choose the image (assumed fixed) 
as the origin of the coordinate system. In this case, the angle 
$\theta$ is still the separation between the image 
and the perturber, but $\phi$ becomes the angle of the 
perturber's position.  One may check that
the above expression remains valid.  The perturbation 
$\delta$ as a function of perturber position is illustrated
in figure ~\ref{geometry}.
Using Equation~(\ref{eq:smallfluxpert}), we 
define the cross section for  magnification
perturbations stronger than $\Delta>0$ as
\begin{equation}
\sigma (\Delta)= \int_{\delta>\Delta} (\theta d\theta d\phi) 
	= \int (\theta d\theta d\phi)\ {\cal H}(\delta - \Delta),
\label{eq:cross-section-definition}
\end{equation}
where $\cal{H}$$(x)$ is the Heaviside step function. 
Of course, if one is interested in perturbations $\delta<0$, 
one ought to integrate over the region 
$\delta<\Delta$ instead.\footnote{We 
rely on context to specify whether the phrase 
``perturbations stronger than $\Delta$'' means 
$|\delta|>\Delta$ regardless of the sign of $\delta$, 
or $\delta<\Delta$ for negative $\Delta$ and 
$\delta>\Delta$ for positive $\Delta$.   For instance,
in this section we will be mostly interested 
in the latter case, but when computing the number
of nonlinear perturbers, 
one is interested in perturbers with 
$|\delta| \gtrsim \Delta \approx 1$ 
regardless of the sign of $\delta$.}
To carry out the integral, one might fix the angle $\phi$, 
find the radial region over which perturbations stronger 
than $\Delta$ exist, and then integrate over all angles.  
Unfortunately, this is not possible in general because the 
equation $\delta(\theta,\phi)=\Delta$ defines the 
$\theta$-coordinate boundary implicitly.
A better approach is to first find the range of radii 
$\theta \in [\tmin,\tmax]$, over which perturbations of size 
$\Delta$ or stronger are possible.  Next, for all radii 
within this interval, find the angular region over which 
perturbers produce perturbations stronger than $\Delta$.  
Then lastly, integrate this angular region over all radii 
to obtain the cross section.  The advantage here is that 
the simple $\phi$ dependence of the perturbation 
$\delta(\theta,\phi)$ allows one to solve explicitly for 
the relevant angular intervals.

%
%

Consider Equation~(\ref{eq:smallfluxpert}).   
As $\theta\rightarrow\infty$, we have $\delta \rightarrow 0$, 
so it follows that if perturbations ever get stronger 
than $\Delta$, there must be a maximum distance 
$\tmax(\Delta)$ at which $\delta=\Delta$ first occurs.  
The quantity $\delta(\theta,\phi)$ is always bounded by 
$2|\mu|((1-\kappa)\delta\kappa \pm \gamma\delta\gamma)$, 
so $\tmax(\Delta)$ is given by the solution to
\begin{equation}
\tmax(\Delta) = \left\{ \begin{array}{lr}
		2|\mu|\left[(1-\kappa)\delta\kappa(\tmax)
       		+ \gamma\delta\gamma(\tmax)\right]
       			& \mbox{if $\Delta>0$} \\
		2|\mu|\left[(1-\kappa)\delta\kappa(\tmax) 
       		- \gamma\delta\gamma(\tmax)\right]
			& \mbox{if $\Delta<0$}
	\end{array} \right .
\label{eq:Theta}
\end{equation}
If perturbations stronger than $\Delta$ are never produced, 
we set $\tmax(\Delta)=0$.  It is important to note that 
even though both $\delta\kappa$ and $\delta\gamma$ are 
monotonic by assumption, $\delta(\theta,\phi)$ need not be.  
If this is the case Equation~(\ref{eq:Theta}), may have 
two roots: $\tmax(\Delta)$ is defined as the larger 
of the two roots, while the second root is the
{\em minimum} radius $\tmin(\Delta)$, 
for which perturbations stronger than $\Delta$ exist.
If Eq.~(\ref{eq:Theta}) has only one or no roots, we 
set $\tmin(\Delta)=0$.  One might expect 
perturbations always to become stronger as 
the perturber approaches the image 
(i.e. that $\delta$ varies monotonically), 
but this is not necessarily so.  
As a counterexample, point masses 
may create negative perturbations on a positive-parity 
image, while singular isothermal sphere (SIS) perturbers 
cannot.  Consider an SIS truncated at some 
finite radius and placed at infinity 
so as to produce a negative perturbation.  
As $\theta$ decreases, 
the perturbation grows more negative, 
until the distance from the image is equal 
to the truncation radius.  At that 
point, the perturbation becomes positive.  
A truncated SIS has a minimum radius 
for which negative perturbations are possible.  

Let us find the angular region over which 
perturbations stronger than $\Delta$ exist 
for a fixed radius $\theta$. 
The boundary points are defined by 
$\delta(\theta,\phi)=\Delta$, with 
a solution for $\phi \in [0,\pi/2]$ given 
by\footnote{It is this step that is not always possible 
to perform when trying to determine the allowed radial 
region for a fixed angle $\phi$.  However, when possible, 
switching the order of integration may prove simpler
\citep[for example, see][]{Keeton03}.}
\begin{equation}
\phim (\theta|\Delta) = 
\frac{1}{2} \cos^{-1} \Big\{\frac{1}{\gamma\delta\gamma(\theta)} 
\Big[ (1-\kappa)\delta\kappa(\theta) 
  - \frac{\Delta}{2|\mu|}\Big]\Big\}.
\label{eq:phi_m}
\end{equation}
Of course, a second solution, $\phi=\pi-\phim$, 
exists in the region $\phi \in [ \pi/2 , \pi ]$. 
In principle, one is also interested in solutions 
with $\phi \in [ \pi, 2\pi ]$, but perturbations 
are symmetric under reflection about the axis $\phi = 0$, 
so the total cross section is double that obtained 
by integrating Eq.~(\ref{eq:cross-section-definition}) 
over the region $\phi\in[0,\pi]$.  
Finally, we can check that $\delta$ always increases with 
$\phi$ from $\phi=\phim$ while $\delta$ decreases with 
$\phi$ from $\phi=\pi-\phim$. The relevant angular interval
for $\delta>\Delta$ with $\Delta>0$ is thus 
$\phi \in [ \phi_m , \pi-\phi_m ]$, 
whereas for $\Delta<0$, the appropriate region is 
$\phi \in [ 0 , \phim ] \cup [ \pi - \phim, \pi ]$.  
Physically, this statement says that perturbers are most 
effective at brightening positive-parity images when 
placed perpendicular to the shear axis $\gamma_1=\gamma$, 
while the opposite is true of negative-parity images. 

One possibility we have not considered is that there is no 
solution to $\delta(\theta,\phi)=\Delta$.  
For $\tmin<\theta<\tmax$, one is guaranteed to have 
perturbations at least as strong as $\Delta$, 
so a lack of solutions implies that at the given radius, 
perturbers at {\em any} angle $\phi$ produce perturbations 
stronger than $\Delta$.  The limit $\delta(\theta,\phi)=\Delta$ 
is never achieved, not because perturbations are not 
strong enough, but because they are always too strong.  
In such cases, the desired angular interval is 
$\phi \in [ 0 , 2\pi ]$, so we define $\phim = 0$ 
for $\Delta>0$, and $\phim = \pi/2$ 
for $\Delta<0$.

We can now compute the magnification perturbation cross section 
by direct integration according to 
Eq.~(\ref{eq:cross-section-definition}).  
This yields 
\begin{equation}
\sigma(\Delta) =
	\left\{ \begin{array}{ll}
	
                \pi\tmax^2 - 
2\int_{\tmin}^{\tmax} (\theta d\theta)\ 2\phi_m(\theta|\Delta) & 
\mbox{if }\ \Delta>0	\\
2\int_{\tmin}^{\tmax} (\theta d\theta)\ 2\phi_m(\theta|\Delta) & 
\mbox{if }\ \Delta<0.
	\end{array} \right.
\label{eq:cross_section}
\end{equation}
Equation~(\ref{eq:cross_section}) is the result we were seeking.  
We summarize our algorithm for computing cross sections as 
follows.

\begin{enumerate}

\item First, given the convergence and shear profiles of a perturber, 
use Eq.~(\ref{eq:Theta}) to find the maximum and minimum radii 
($\tmin$ and $\tmax$) such that perturbations $\delta$ 
stronger than $\Delta$ exist.

\item Second, define the function $\phim(\theta|\Delta)$ via 
equation ~(\ref{eq:phi_m}). 

\item Third, use Eq.~(\ref{eq:cross_section}) to compute 
	the total magnification perturbation cross section.

\end{enumerate}
In appendices \ref{ptmass} and \ref{sis} we 
illustrate this formalism using point mass and
singular isothermal sphere profiles as simple examples.  
This analysis reproduces the previous results of 
\citet{MaoSchneider98} and \citet{Keeton03} in these 
two cases respectively.

%
%
\section{The Effects of Multiple Perturbers: 
The Average Perturbation and Its Variance}
\label{section_multiple_perturbers}

In the previous section, we studied the effect of a single 
perturber on an image and computed the cross sections for 
perturbations of a given amplitude.	
In this section, we extend these results and 
consider magnification perturbations due 
to an ensemble of perturbers.  In particular, we compute the 
average perturbation $\avg{\dt}$ and its variance in the limit 
that large magnification perturbations are unlikely and 
assuming that no individual perturber creates a perturbation 
larger than some linear cutoff $\dnl$. 

Consider the magnification perturbation due to an ensemble of 
perturbers. Rather than adding these perturbers artificially 
and thereby increasing the mass of the lens, 
we assume the perturbers constitute a redistribution of 
some fraction of the mass of the macrolens.  
This is equivalent to adding substructure 
while at the same time adding a negative mass component that 
traces the macrolens' mass distribution.  
Thus, we first treat the problem of simply
adding substructure artificially, and at the end we include
the effect of mass conservation by adding the magnification
perturbation due to a hypothetical negative mass component.  

In our linear approximation, the total substructure 
magnification perturbation is the sum of the individual 
perturbations 
\begin{equation}
\dt (N) = 2|\mu|\left[ (1-\kappa)\sum_{i=1}^N \delta\kappa_i 
           - \gamma\sum_{i=1}^N\delta\gamma_i\cos(2\phi_i)\right]
        = \sum_{i=1}^N \delta_i. 
\label{eq:totpert}
\end{equation}
In Eq.~(\ref{eq:totpert}), $\delta_i$, $\delta \kappa_i$, and 
$\delta \gamma_i$ represent the magnification perturbation, 
convergence perturbation, and shear perturbation of the $i$th 
perturber respectively.  
This relation should hold provided that the 
position perturbation $\delta \balpha$ is negligible. 
Unlike the case of a single perturber, in the case 
of multiple perturbers, we do not have 
a ``rule-of-thumb'' condition for when this 
approximation holds.  However, 
we na{\"{\i}}vely expect that the approximation will 
be good when magnification perturbations are small.  
In particular, oppositely-positioned perturbers give no net 
astrometric perturbation, but no analogous symmetry exists for 
magnification perturbations so long as $\delta\kappa>0$ (that is, 
the perturber represents an increase in the surface density).   
Therefore, we expect position perturbations to grow no faster
than magnification perturbations.  Henceforth, we 
assume Eq.~(\ref{eq:totpert}) is valid 
(astrometric perturbations are negligible), 
and expect such an approximation to be justified if 
the net magnification perturbation is small, as 
in Eq.~(\ref{eq:dsmall}).

Given a perturber with a uniform probability of being 
in some finite region of space $\cal{R}$, about an image, 
the probability distribution $\rho(\delta)$, for a 
magnification perturbation in the range $\delta$ to 
$\delta$+$d\delta$ is given by
\begin{equation}
\rho(\delta)d\delta = \frac{1}{A(\cal{R})} 
\int_{\cal{R}} d^2\btheta\
	\dirac\left[\delta(\theta,\phi)-\delta\right] 
	= \frac{1}{A(\cal{R})} 
          \left|\frac{d\sigma}{d\delta} \right| d\delta,
\label{eq:prob_dist}
\end{equation}
where $\sigma(\delta)$ is the cross section for the perturber 
to produce a perturbation stronger than $\delta$,  
$A(\cal{R})$ is the area of the region $\cal{R}$, and 
$\dirac[x]$ is the Dirac delta function of $x$.  
Though a uniform probability distribution 
for the perturber is unphysical, 
it is a reasonable approximation provided that 
the region over which perturbers can sizeably 
affect image magnifications is small compared 
to the length scale over which the {\em projected} 
number density of substructures varies.

We compute the various moments of the distribution 
$\avg{\delta^n}$, by defining the region $\cal{R}$ 
as all points $(\theta,\phi)$ such that 
$\dmin < | \delta(\theta,\phi) | < \dnl$,  
In other words, we assume that the perturber 
creates a perturbation stronger than some minimum perturbation 
$\dmin$, but weaker than the cut off $\dnl$.  
Taking expectation values, and using the fact that 
$d\sigma(\delta)/d\delta < 0$ for $\delta > 0$ and 
$d\sigma(\delta)/d\delta > 0$ for $\delta < 0$ 
to integrate by parts, we obtain
\begin{eqnarray}
\avg{\delta^n} & = & \frac{1}{A(\cal{R})} 
\Bigg[ -\dnl^n y_n(\dnl) + \dmin^n y_n(\dmin) 
\white \bigg] \nonumber \\
 & + & \white \Bigg[ \black 
n \int_{\dmin}^{\dnl} d\delta\ 
\delta^{n-1}y_n(\delta) \Bigg],
\label{eq:singlepert}
\end{eqnarray}
where we have defined the function 
$y_n(\delta) = \sigma(\delta) + (-1)^n\sigma(-\delta)$.  

Consider an ensemble of identical perturbers 
characterized by a parameter vector ${\bf p}$ that 
specifies their projected mass density profiles 
(e.g., mass for point masses, or velocity dispersion for SIS 
perturbers).  The expectation value of the 
total perturbation is 
\begin{eqnarray}
\avg{\dt} & = & \sum_{N=0}^\infty  P(N) 
\int\left(\prod_{i=1}^{N}\rho(\delta_i)d\delta_i\right)
\sum_{i=1}^N \delta_i({\bf p}) \\ 
	& = & \avg{N}\avg{\delta|{\bf p}},
\label{eq:avgpert1}
\end{eqnarray}
where the quantity $\avg{\delta|{\bf p}}$ is the expectation 
value of $\delta$ due to an individual perturber with 
parameters ${\bf p}$, and $P(N)$ is 
the probability of finding $N$ perturbers 
in the region $\dmin \le |\delta| \le \dnl$.
A similar computation shows that 
the variance 
$\Var{\dt} = \avg{\dt^2} - \avg{\dt}^2$
is given by
\begin{eqnarray}
\Var{\dt} & = & \avg{N} \avg{\delta^2|{\bf p}}
             + [\avg{N^2} - \avg{N}^2 - \avg{N}] 
              \avg{\delta|{\bf p}}^2 \nonumber \\
           & = & \avg{N} \avg{\delta^2|{\bf p}}, 
\label{eq:varpert1}
\end{eqnarray}
where, similar to the previous equation, 
$\avg{\delta^2|{\bf p}}$ is the expectation value 
of $\delta^2$ due to each individual perturber. 
To derive the second equation, we assumed that 
$P(N)$ is Poisson, such that 
$\Var{N} = \avg{N}$, with no 
correlation in the spatial distribution of 
perturbers.\footnote{Strictly speaking, one would expect 
the perturbers to be correlated.  However, 
if the surface density of perturbers is sufficiently 
low, such correlations should be a small effect.}

We can extend this discussion to an ensemble of 
perturbers with different properties.  
Let $ds/d{\bf p}$ be the number density of 
perturbers with parameters ${\bf p}$ within 
some infinitesimal volume of parameter space $d{\bf p}$.  
The expected number of perturbers becomes 
$\avg{N} = \dl^2 A({\cal{R}}) ds({\bf p})/d{\bf p}$, 
where $\dl$ is the angular diameter distance to the lens.  
Placing this into our expressions for the 
average perturbation and summing over 
all perturbers (all parameters ${\bf p}$), 
we obtain 
\begin{eqnarray}
\avg{\dt} & = & \int d{\bf p}\ \frac{ds}{d{\bf p}} \dl^2 
\Bigg[ - \dnl y_1(\dnl) + \dmin y_1(\dmin) \white \Bigg] 
 \nonumber \\
	   & + & \white \Bigg[ \black  
                  \int_{\dmin}^{\dnl} d\delta\ y_1(\delta) \Bigg],
\label{eq:avgdt}
\end{eqnarray}
where $\dnl$ and $\dmin$ are the maximum and 
minimum individual perturbations respectively.  
The total perturbation is obtained by taking 
the limit $\dmin \rightarrow 0$.  
Note that it is possible for the integral over 
$\delta$ to diverge (see SIS perturbers in appendix \ref{sis}).  
Likewise, the variance of the total perturbation is
\begin{eqnarray}
\Var{\dt} & = & \int d{\bf p}\ \frac{ds}{d{\bf p}}\dl^2 
 \Bigg[ -\dnl^2 y_2(\dnl) + \dmin^2 y_2(\dmin) 
  \white \Bigg] \nonumber \\
           & + & \white \Bigg[ \black 
     2\int_{\dmin}^{\dnl} d\delta\ \delta y_2(\delta) \Bigg].
\label{eq:vardt}
\end{eqnarray}
Including mass conservation (the negative mass component 
mentioned above) is now trivial.  If the mass density in
substructures  is $\fsat\Sc/2$, 
the net perturbation becomes
\begin{equation}
\avg{\dt} = \avg{\dt}_{\mathrm{pert}} 
            - |\mu| (1 - \kappa ) \fsat
\label{eq:negmass}
\end{equation}
where the perturber contribution $\avg{\dt}_{\mathrm{pert}}$ 
is given by Equation~(\ref{eq:avgdt}) above and we used the
fact that the convergence perturbation from the negative mass
component is $ \delta \kappa = - \fsat/2 $.  
This represents a constant contribution, 
so the variance remains unchanged.
In Appendices~\ref{ptmass}~and~\ref{sis}, 
we consider the average perturbation 
$\avg{\dt}$ and its variance $\Var{\dt}$ 
for the case of point mass and 
SIS perturbers respectively.

%
%
\section{Magnification Perturbations by Halo Substructures}
\label{substruct}

In this section we apply our formalism to the problem of 
magnification perturbations by substructures 
within the context of a simple, but realistic, model 
of CDM substructure that includes a mass spectrum of 
perturbers and tidal truncation of the substructure profiles.
We find that such perturbations are local in that 
the assumption of a constant surface density of perturbers 
is valid.   We find that while the most massive perturbers
dominate the total magnification perturbation,
the latter has significant contributions from a large
range of substructure masses and these contributions 
depend upon the shape of the substructure mass spectrum.  
In what follows, we ignore the negative mass
component perturbation in \S~\ref{sub:local}, 
\S~\ref{sub:mspectrum}, 
and \S~\ref{truncating_mass_spectrum},
which deal with the details of computing the granular
contribution to the magnification perturbation.  
We include the negative mass contribution in 
\S~\ref{predictions}, where we illustrate 
the expectation values for the specific models.

%
%
\subsection{Substructure Perturbations Are Local}
\label{sub:local}

Following the analysis of \citet{DalalKochanek02a}, 
we adopt a simplified model for substructures in which
perturbers are characterized by so-called 
pseudo-Jaffe mass density profiles 
\citep[see][]{munoz_etal01}.  
The three-dimensional density distribution is 
\begin{equation}
\label{eq:jaffe3d}
\rho(r) \propto \frac{1}{r^2 (r^2 + a^2)},
\end{equation}
where the parameter $a$ is an effective tidal radius 
that should physically be present for all substructures 
and serves to localize the total mass of the 
perturber.\footnote{We note that while a pseudo-Jaffe profile 
is not the most accurate treatment for the density profiles of 
dark matter substructures observed in simulations, it is 
qualitatively correct in that the matter density quickly drops 
beyond a truncation radius $a$.  Due to its simplicity, 
and because it allows us to better relate to previous work which uses 
this or related profiles, we have opted to model substructures 
as pseudo-Jaffe profiles.  Adiabatic contraction is one motivation 
for using steeper inner profiles, but this is likely to be irrelevant 
for the majority of small, subhalos that do not host luminous galaxies.  
The formalism that we present can be adapted to accommodate more general 
subhalo density profiles.}  
The two-dimensional convergence and shear profiles for 
the pseudo-Jaffe model are given by 
\begin{equation}
\delta\kappa = \tb \left(\frac{1}{2\theta}-\frac{1}{2\xi} \right),
\label{eq:pseudoJaffek}
\end{equation}
and
\begin{equation}
\label{eq:pseudoJaffeg}
\delta\gamma = \frac{\tb}{2\theta} + \frac{\tb}{2\xi} 
- \frac{\tb a}{\theta^2} \left(\frac{\xi}{a}-1\right),
\end{equation}
where $\tb$ is defined by the combination 
$b/\tb=1+a/b-(1+(a/b)^2)^{1/2}$, 
the quantity $b$ is the Einstein radius of the profile, 
and $\xi \equiv \sqrt{\theta^2+a^2}$.  
When $a \rightarrow \infty$, the 
profiles approach that of an SIS perturber with an 
Einstein radius $b = \tb$, while 
in the limit $ a \rightarrow 0$ with $a\tb$ constant, 
the perturber corresponds to a point mass with Einstein 
radius $b = \sqrt{a\tb}$.  The total mass of the perturber 
is given by $m = \pi a \tb \Sc \dl^2$.  

%
%
\begin{figure}[t]
\epsscale{1.0}
\plotone{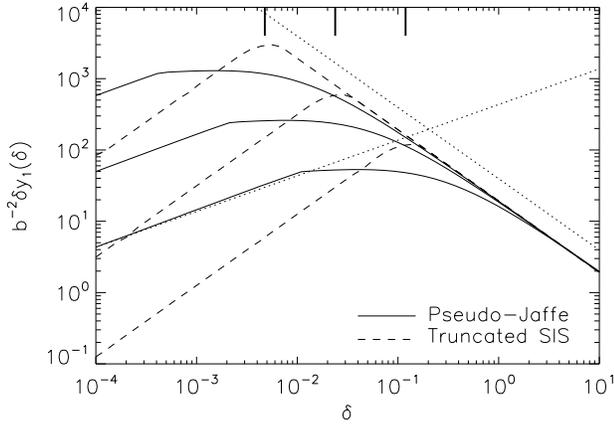}
\caption{
The contribution per logarithmic interval to the 
total perturbation due to perturbers of strength $\delta$. 
The quantity $\delta y_1(\delta)$ (see text) must be 
multiplied by a number density to get a perturbation, 
so we plot the dimensionless quantity 
$b^{-2}\delta y_1(\delta)$ where $b$ is the perturber's Einstein
radius.
All of the curves assume a macromodel with 
$\kappa=0.5$ and $\gamma=0.2$.  From top to bottom the 
three {\em solid} and {\em dashed} lines correspond to 
perturbers with  $a/b = 500, 100, 20$, 
where $a$ is the substructure tidal radius.
The {\em solid} lines show the relation for 
pseudo-Jaffe profiles [cf. Eq.~(\ref{eq:jaffe3d}), 
Eq.~(\ref{eq:pseudoJaffek}), Eq.~(\ref{eq:pseudoJaffeg})]. 
The {\em dashed} lines are truncated SIS profiles.
The falling {\em dotted} line shows the expected scaling 
for SIS perturbers, $\delta y_1(\delta) \propto 1/\delta$, 
while the rising {\em dotted} line shows the
expected scaling in the limit $\delta \rightarrow 0$.  
The broad peak for the pseudo-Jaffe profiles reflects
the smooth change from 
$\delta y_1 \propto \delta^{-1}$ 
to $\propto \delta^{1/3}$.  
This transitions to $\propto \delta^{1/2}$ 
when perturbations become shear dominated 
(sharp break).  
The peak scales $\deltap = |\mu| (1-\kappa) b / a$ 
are marked with {\em short, vertical} lines at the 
top of the Figure.
}
\label{y1}
\end{figure}

First, consider the average perturbation $\avg{\dt}$, 
for a positive-parity image.  
Perturbers well within a tidal radius from the image 
act as SIS ($\delta\kappa = \delta\gamma\sim1/\theta$), 
so each individual perturbation $\delta$ scales as
$\delta \sim \delta \kappa \sim 1/\theta$, 
while the number of perturbers per logarithmic 
interval grows as $\sim \theta^2$.  
As a result, the 
contribution to $\dt$ scales as 
$\sim \theta \sim 1/\delta$, meaning that it 
{\em increases} with {\em decreasing} individual 
perturber strength, until a peak at 
$\deltap \approx |\mu| (1-\kappa) b/a$, which is the 
typical perturbation from substructure 
{\em one tidal radius away from the image}.  

At large distances the behavior is slightly more 
complicated.  One might na{\"{\i}}vely expect point mass 
behavior to dominate, but the average perturbation 
due to point masses is zero.  
A nonzero contribution to $\dt$ arises from the small, 
but nonzero, convergence $\delta \kappa \sim 1/\theta^3$ 
(valid for $\theta \gg a$).  
Therefore, the contribution from perturbers a 
distance $\theta$ away scales as 
$\sim \theta^2\delta\kappa \sim 1/\theta$, 
which, for convergence-dominated perturbations, 
is a $\sim \delta^{1/3}$ scaling.  
In the limit $\delta \rightarrow 0$, 
all perturbations eventually become shear dominated 
with the individual perturber strength 
$\delta \sim 1/\theta^2$, so the net contribution 
to $\dt$ scales as $\sim 1/\theta \sim \delta^{1/2}$.

We illustrate this behavior explicitly in Figure~\ref{y1}, 
where we assume typical macromodel parameters 
$\kappa = 0.5$ and $\gamma = 0.2$.
For comparison, we also present the case for truncated 
SIS perturbers.  At first glance,
the curve for truncated SIS perturbers seems puzzling 
because the curve does not drop off to zero
at the tidal radii ($\delta < \deltap$).  
The reason for this behavior is that we have been 
somewhat loose in our interpretation of 
$y_1(\delta)\delta$.  In particular, 
the full contribution to $\dt$ per logarithmic 
interval is 
$\delta \rho( \delta ) = \delta dy_1/d\delta$, 
which is zero in the region $\delta < \deltap$.
To derive Eq.~(\ref{eq:avgdt}), 
we integrated $\delta dy_1(\delta)/d\delta$ by parts.  
This {\em halves} the contribution from the region 
$\delta > \deltap$, while introducing a 
symmetrical contribution in the region 
to the left of the peak.

\begin{figure}[t]
\epsscale{1.0}
\plotone{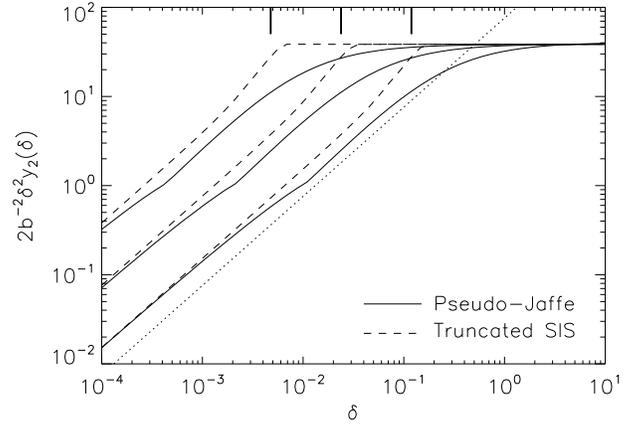}
\caption{
The contributions per logarithmic interval to the 
variance of magnification perturbations.  
The lines are the same as in Figure~\ref{y1}, 
but the vertical axis now represents the 
contribution to the variance due to perturbers of 
strength $\delta$ in the dimensionless form 
$b^{-2}\delta^2y_2(\delta)$.  
Again, the macromodel has $\kappa=0.5$ and $\gamma=0.2$, 
and the substructure Einstein $(b)$ and tidal $(a)$ radii satisfy
$a/b=500,100,20$ from left to right.  
The {\em dotted} line is the expected
scaling for point masses, 
$\delta^2y_2(\delta)\propto\delta$, 
and the {\em short, vertical} 
lines at the top mark the peak scale 
$\deltap = |\mu| (1-\kappa) b/a$.
}
\label{y2}
\end{figure}

Analogous arguments hold for the variance of the 
magnification perturbations, $\Var{\dt}$.  
Perturbers within a tidal radius contribute 
$\sim \theta^2 d\ln(\theta) \delta^2 \sim d\ln(\delta)$
to the variance, so all perturbers of strength 
$\delta < \deltap$ contribute at comparable levels 
to $\Var{\dt}$.  
Perturbers far away act as point masses 
and contribute an amount that scales as 
$\sim \theta^2 \delta^2 \sim \delta$, 
where we used $\delta \sim 1/\theta^2$ for point masses.  
Finally, there is a small transition region
between the peak and point-mass behavior where 
perturbations are convergence dominated with 
$\delta \kappa \sim 1/\theta^3$, 
corresponding to a $\sim \delta^{4/3}$ scaling.
We show the contribution to the variance from strength 
$\delta$ perturbers in Figure~(\ref{y2}).

%
%

The above results have some interesting implications.  
In particular, we have demonstrated that:
(1) the average perturbation is dominated by perturbers 
that are about one tidal radius away from the image; 
and (2) the variance of the perturbations weighs all 
perturbers within a tidal radius equally, 
with perturbers further away contributing relatively little.  
In other words, magnification perturbations are effectively 
``local" to within the typical tidal radii of the perturbers.  
How local is this?  Consider a perturber a distance $d$, 
from the center of the macrolens potential. 
Then $d = \sqrt{r^2 + z^2}$, where $r$ is its 
projected distance onto the lens plane and 
$z$ is the separation along the line of sight.  
Assuming, for simplicity, truncated SIS profiles for 
both host and perturber, we can approximate the tidal 
radius $a(r,z)$, by setting 
$\rho_{perturber}(a) = \rho_{host}(d)$.  
We are generally interested in the 
properties of substructures at a projected distance 
roughly equal to the Einstein radius of the macrolens, 
$r = b_H$.  
We find $(b/a)^2 = bb_H/(b_H^2+z^2)$.
Assuming, for simplicity, that substructures trace the 
smooth component, we can obtain an estimate for the average 
value of $b/a$ along the line of sight to the lens by 
integrating along the line-of-sight distance, $z$.  
This yields 
\begin{equation}
\avg{b/a} \approx \frac{1}{\lambda_a} (2/\pi)(b/b_H)^{1/2},
\label{eq:tidrad}
\end{equation}  
with $\lambda_a = 1$.  For comparison, 
\citet{DalalKochanek02a} used
$\lambda_a=2/\pi \simeq 0.64$.  
We introduced the parameter $\lambda_a$, 
as a way to change the effective tidal radii 
in our model.  Larger $a$ corresponds to 
systematically less centrally-concentrated 
density profiles.  \citet{KochanekDalal04} assumed 
$\lambda_a=2/\pi$ in their analysis.
For $b/b_H \sim 10^{-3} - 10^{-2}$ 
(e.g., milliarcsecond perturbers), 
we obtain $a/b_H \approx 5\%-15\%$.  
This distance is small enough that 
perturbations may be considered local 
in the sense that the two-dimensional 
surface density of perturbers does not 
vary much on this length scale 
\citep[cf.][]{zentner_bullock03,Maoetal04,
diemand_etal04,nagai_kravtsov05}, 
but large enough that magnification perturbations 
to nearby pairs of images may be correlated, 
especially if the substructure profiles are  
less centrally concentrated than we have assumed 
($\lambda_a >1$).  

%
%
\subsection{Dependence of Magnification Perturbations 
on the Substructure Mass Spectrum}
\label{sub:mspectrum}

We now investigate the dependence of magnification 
perturbations on the mass spectrum of the 
substructure perturbers.  Consider Equation~(\ref{eq:avgdt}) 
for the average perturbation.  
We approximate the mass spectrum as a power law 
$ds/dm \propto m^{\alpha}$, truncated at some maximum mass 
$\mmax$, where $ds/dm$ is the number density of substructures 
in the mass range from $m$ to $m + dm$.  
The slope $\alpha$ has been determined by high-resolution, 
cosmological $N$-body simulations, which generally yield  
$\alpha \approx -1.8\ \pm 0.1$
\citep[e.g.,][]{Klypinetal99,Mooreetal99,
delucia_etal04,diemand_etal04,gao_etal04a}.
The surface mass fraction $d\fsat$ of perturbers 
of mass $m$ is 
$(1/2)\Sc d\fsat = d\Sigmas = mds$, 
so that the mass fraction varies as 
$d\fsat/dm \propto m^{\alpha+1}$.  
Here, we have chosen to follow 
\citet{DalalKochanek02a} 
and define the ``mass fraction'' $\fsat$ 
relative to half of the critical density, 
$\fsat = 2\Sigmas/\Sc$.  
This choice is motivated by the fact that
all images in a lens will probe 
similar radii in the host halo, 
and thus similar substructure densities, 
whereas the macromodel convergence may vary 
more strongly from image to image.
Note that the quantity $mds/dm$ is precisely the 
number of perturbers of mass $m$ 
per logarithmic mass interval, so that 
Eq.~(\ref{eq:avgdt}) can be recast as
\begin{eqnarray}
\avg{\dt} & \approx  & \half \int_0^{\mmax} 
\frac{dm}{m} \frac{d\fsat}{dm}\Sc\dl^2
\int_{\dmin}^{\dnl} d\delta\ y_1(\delta) \nonumber \\
 & \approx &  \half(\alpha+2)\fsat\int_0^{\mmax} \frac{dm}{m} 
	\left(\frac{m}{\mmax}\right)^{\alpha+2} \nonumber \\
 & \times & \left[\frac{\Sc\dl^2}{m} 
	\int_{\dmin}^{\dnl} d\delta\ y_1(\delta)\right], 
\label{eq:perteq}
\end{eqnarray}
where we have ignored the contribution from 
the boundary terms by taking the limits 
$\dmin \rightarrow 0$ and $\dnl \rightarrow \infty$, 
and we have normalized the substructure fraction 
as $\fsat = \int_{0}^{\mmax} dm\ d\fsat/dm$, 
neglecting a possible low-mass cutoff by 
assuming that $\alpha > -2$.  
The mass $\mmax$ represents the mass 
scale of the most massive substructures 
contributing to the magnification perturbations.  
Below, we argue that this is given roughly by 
$\mmax/M_E \sim 1 \%$, where $M_E$ is the mass of the 
macrolens contained within an Einstein radius in 
projection.

%
%
\begin{figure}[t]
\epsscale{1.0}
\plotone{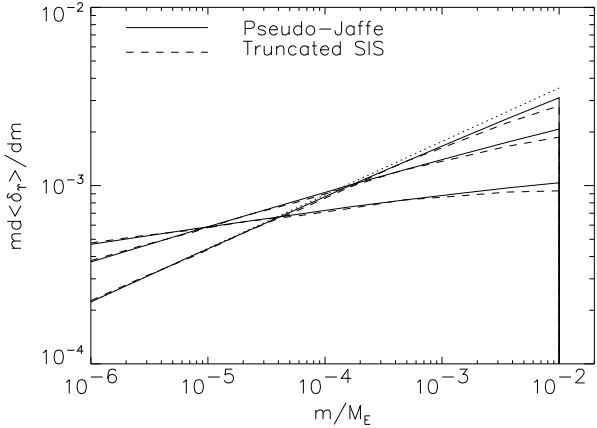}
\caption{
The average magnification perturbation per logarithmic 
interval in substructure mass $m$.  
Masses are expressed in units of the smooth lens mass 
$M_E$ contained within the Einstein ring, and the most
massive substructures considered are 
$\mmax/M_E = 1\%$ (see text).
A substructure mass fraction of 
$\fsat = 5 \times 10^{-3}$ with a power-law 
spectrum of index $\alpha$ was assumed. 
The macromodel parameters are 
$\kappa = 0.5$ and $\gamma = 0.2$. 
The {\em solid} lines are for the 
pseudo-Jaffe profiles, and the {\em dashed} 
are for truncated SIS profiles.
From top to bottom at right, the value of 
$\alpha$ is $\alpha=-1.7,-1.8,$ and $-1.9$.
The {\em dotted} line shows the expected 
$m^{2+\alpha}$ scaling for $\alpha = -1.7$.
}
\label{sy1}
\end{figure}

With these assumptions we can estimate how the 
integrand in Eq.~(\ref{eq:perteq}) scales with mass.  
The perturbation by mass $m$ perturbers
is dominated by substructures a tidal radius 
away from the image and scales as 
$\sim a^2 ( |\mu| b/a ) \sim ab \sim m$.
Therefore, the term in square brackets is roughly constant, 
and the perturbation per logarithmic mass interval varies 
as $\sim m^2 ds \sim m d\fsat \sim m^{\alpha + 2}$ or as 
$\sim m^{0.1-0.3}$ for 
$\alpha = -1.7 \  \mathrm{to}\  -1.9$.  
This dependence is rather weak.  
We show the $m^{2+\alpha}$ scaling explicitly in 
Figure~\ref{sy1}, 
after enforcing a linear cutoff $\dnl = 1$ 
(replacing the boundary term that we neglected earlier).  
We show curves for both pseudo-Jaffe and 
truncated SIS profiles.  
As is evident in the Fig.~\ref{sy1}, 
the results are nearly identical 
for these two cases.  
Finally, the term in the square brackets 
in Equation~(\ref{eq:perteq}) is, 
to a good approximation, a constant function 
of perturber mass, so that we can write
\begin{equation}
\avg{\dt} \approx \half\fsat 
	\left[\frac{\Sc\dl^2}{m} 
\int_{\dmin}^{\dnl} d\delta \  y_1(\delta)\right]_{m=\mmax}.
\label{eq:perteq2}
\end{equation}
The error introduced by making this approximation is 
$\approx 5\%$ and is nearly independent 
of the exact value of $\mmax$ so long as the 
peak scale $\deltap < \dnl$, where we reiterate that 
$\dnl$ is a linear cutoff scale.

\begin{figure}[t]
\epsscale{1.0}
\plotone{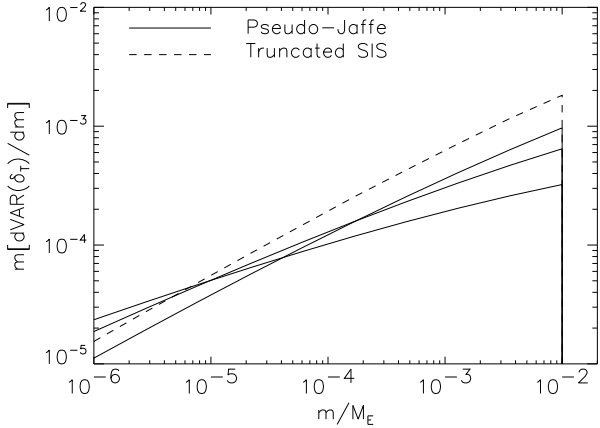}
\caption{
The variance of the magnification perturbation
per logarithmic mass interval. 
A substructure mass fraction of $\fsat = 0.5\%$ with
a power-law spectrum of slope $\alpha$ was assumed.
The linear cutoff $\dnl$ is set to unity and
the macromodel parameters are $\kappa=0.5$ and 
$\gamma=0.2$.  The {\em solid} lines are for the 
pseudo-Jaffe profile and the value of $\alpha$
is, from top to bottom at right, 
$\alpha=-1.7,-1.8$, and $-1.9$.  The
{\em dashed} line is for a truncated SIS
profile with $\alpha=-1.7$.
}
\label{sy2}
\end{figure}

As before, we can construct a similar argument for the 
variance of the magnification perturbations.  
In this case, all perturbers within a tidal radius 
contribute equally to the variance, so a linear cutoff 
$\dnl$, needs to be introduced.  
This constant contribution scales as 
$\sim a^2(b/a)^2 \sim b^2 \sim m^{4/3}$, 
and is integrated between the peak scale 
$\deltap$ and the linear cutoff $\dnl$.  
The contribution per logarithmic mass interval is thus 
$\sim (m d\fsat) m^{4/3} \ln(\Delta/\deltap) 
\sim m^{\alpha + 7/3} \ln(\Delta/\deltap)$.  
For $\alpha=-1.8$, this is 
$\sim m^{8/15} \ln(\Delta/\deltap)$.  
As before, we can use our understanding
of this behavior to provide a simple estimate 
of the variance,
\begin{eqnarray}
\label{eq:vardt2}
\Var{\dt} & \approx & \half(\alpha+2)\fsat
	\left[\frac{2\Sc\dl^2}{m} \int_0^{\dnl} 
	d\delta\ \delta y_2(\delta)\right]_{m=m_c} \nonumber \\
 & \times & \int_0^{\mmax}\frac{dm}{m} 
	\left(\frac{m}{\mmax}\right)^{ 2 + \alpha+\frac{1}{3}}
	\Bigg[ \frac{\ln\left(\Delta/\deltap(m)\right)}
	{\ln\left(\Delta/\deltap(\mmax)\right)} \Bigg].
\end{eqnarray}
This approximation is valid to about $\approx 5\%$ and, 
once again, it breaks down as $\deltap$ approaches 
$\dnl = 1$.  
We illustrate these results in Figure~\ref{sy2}, 
and we note particularly the flatness of the variance 
per logarithmic mass interval, which is
shallower than $d \ln [\Var{dt}] / d \ln m \lsim 1/2$.  
Note that the expression above differs from that
of identical mass $\mmax$ perturbers, 
demonstrating that the statistical properties of 
magnification perturbations by substructures
depend on the mass spectrum of the perturbers.  
This is not surprising as 
such a dependence has been noted in the 
case of microlensing 
\citep[e.g.,][]{Schechteretal04}.
Finally, from Figure~\ref{sy2}, 
we see that SIS and pseudo-Jaffe profiles 
lead to different variances at a fixed mass fraction, 
implying that the variance of the magnification 
perturbation is sensitive to the substructure 
mass density profile.

%
%

%
%
\subsection{Truncating the Substructure Mass Spectrum}
\label{truncating_mass_spectrum}

It is interesting to consider what the appropriate values 
for the effective minimum and maximum masses, 
$\mmax$ and $\mmin$, are.  
We begin by considering the relatively simpler 
problem of the low-mass cutoff, $\mmin$.  
In general, the magnification perturbation 
due to a perturber smaller than the physical 
extent of the source region is small 
\cite[e.g.,][]{Schneideretal,DoblerKeeton05}.  
The average perturbation is dominated by perturbers 
a distance $\simeq a$ from the source, 
where $a$ is the tidal radius of the substructure, 
so the point source approximation should be valid 
so long as the characteristic size of the source $R$ 
is smaller than $a$.  The minimum substructure mass 
can therefore be obtained by setting $a \approx R$.  
Taking $R \sim 10\ \pc$, as appropriate for quasar 
radio-emission regions, we find
$\mmin/M_E \gtrsim \lambda_a^{-2}10^{-9}$, 
small enough to neglect for our present 
purposes\footnote{One could require instead $b \approx R$, 
which is appropriate in the limit that individual 
perturbations dominate the net magnification perturbation.  
This gives $\mmin/M_E \gtrsim \lambda_a 10^{-5}$.  
In our model, the number of nonlinear perturbers of 
mass $m$ scales as 
$m^{\alpha + 1}b^2 \sim m^{\alpha + 7/3} \sim m^{1/2}$, 
implying that the number of nonlinear perturbers increases 
with mass.  At most, or order one nonlinear 
perturbers will be present, and 
these will all be massive.  
Hence, the appropriate limit for 
$\mmin$ is that of linear perturbers
obtained by setting $R \approx a$ as above.}.

Estimating the appropriate maximum effective 
substructure mass $\mmax$, is a considerably 
more difficult problem.  In the context of CDM, 
the most massive substructures within a 
halo have masses that are typically of order 
$\mmax \sim 10^{-2} \Mvir$, where $\Mvir$ is the 
virial mass of the halo 
\citep[e.g.,][]{Klypinetal99, Mooreetal99, 
ghigna_etal00, delucia_etal04, 
diemand_etal04, gao_etal04a}.  
For a halo with $\Mvir = 10^{13} \msun$, 
this corresponds to $m \sim 10^{11} \msun$, 
which is comparable to the mass of the host halo 
contained within a cylinder of radius equal to the
Einstein radius of the lens.
Clearly, if such a substructure were to fall near the 
Einstein radius of the host halo, all images would be 
strongly perturbed and the lens macromodel would be 
significantly altered 
\citep[see, e.g.,][and references therein]{CohnKochanek04}.  
In the most na{\"{\i}}ve view, one could imagine that 
such effects are either entirely degenerate with 
the macromodel, or else explicitly including them 
when the substructures can be optically identified.
In this na{\"{\i}}ve picture, 
magnification perturbations come about only 
through substructures that do not affect the macromodel.  
Taking the primary macromodel property to 
be the Einstein radius, one can demand that 
$\mmax \ll M_E \propto b_H^2$, 
so one may expect $\mmax/M_E = 1\%$ 
to be a reasonable cutoff mass.  

A more pragmatic self-consistency 
argument can also be made.  We have
assumed in our formalism that we can neglect 
astrometric perturbations relative to the magnification 
perturbations, so we should include only perturbers for 
which the characteristic perturbation scale $\deltap$ 
is linear.  We can then define $\mmax$ 
by setting $\deltap(\mmax) =1 $, which leads to
$\mmax \sim \lambda_a^4 (\pi/2)^4 |\mu|^{-3}$.  
For $\lambda_a = 1$, $|\mu|=10$, and $\deltap=1$, 
we get $\mmax/M_E \sim 0.6\%$, though it is clear that 
this value depends upon the various parameters.  
However, it is important to note that 
less centrally-concentrated profiles ($\lambda_a>1$) 
allow for significantly larger mass cutoffs, 
implying that linear theory becomes a better 
approximation.  In what follows, we truncate 
the mass spectrum at $\mmax/M_E = 1\%$, 
regardless of the macromodel parameters.

It should be clear from these arguments 
that the truncation scale $\mmax$ carries
with it a some uncertainty.  
In general, any given macromodel should be 
able to partly, but not fully, 
compensate for perturbations due to 
very massive substructures,
in which case there will be 
an additional contribution to the 
observed magnification perturbations 
that is not being included by introducing a 
cutoff $\mmax/M_E = 1\%$.  
Estimating the degree to which the effect 
of massive substructures can be absorbed 
within the macromodel is beyond the scope of
this work, but we explore this in a forthcoming 
companion paper (Chen \& Rozo 2005, In preparation).

%
%
\subsection{General Predictions}
\label{predictions}

%
%
\begin{figure}[t]
\epsscale{1.0}
\plotone{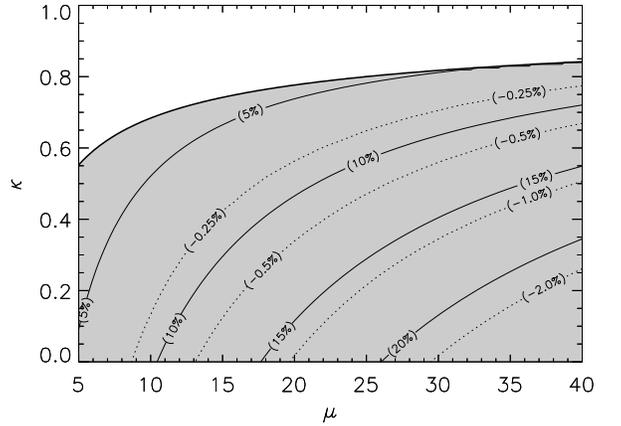}
\caption{
The average and rms values for the total magnification 
perturbation due to linear perturbers of an image with 
macromodel magnification $\mu$ and convergence $\kappa$.
{\em Solid} contours give the rms values while
{\em dotted} contours give the average perturbation.
Lens and source redshifts of $\zl = 0.3$
and $\zs = 1$ respectively, were assumed, 
and a high-mass cutoff 
$\mmax / M_E = 1\%$, where $M_E$ is the mass of the 
macrolens projected to within an Einstein radius. The
tidal radii were set by $\lambda_a = 4$.  
The substructure mass fraction is $\fsat = 0.5\%$,
and mass
conservation is enforced through a negative mass component
of surface density $-\kappa_s=2\Sc\fsat$.  
The unshaded region is unphysical as it 
corresponds to $\gamma < 0$.
The rms scatter is comparable to that shown in 
Figure 3 of \citet[][]{KochanekDalal04}. 
}
\label{moments}
\end{figure}

We can use this formalism to make some general 
predictions regarding magnification perturbations 
due to halo substructure.  
Figure~\ref{moments} illustrates predictions for 
the average and rms value of the 
total magnification perturbation 
in the case of positive-parity images.
As an illustrative example, 
we have assumed a source redshift of $\zs = 1$ 
and a lens redshift of $\zl = 0.3$.  
We have also assumed a high-mass cutoff 
$\mmax / M_E = 1\%$, a substructure mass fraction 
$\fsat=2\Sigma_s/\Sc = 0.5\%$, 
a tidal radius given by $\lambda_a = 4$, 
and a typical host Einstein radius of $b_H = 1"$.  
Although the average magnification 
perturbations are very small ($\lesssim 1\%$), 
the rms fluctuations are quite large $(\drms \gtrsim 10\%)$, 
and hence large magnification perturbations should be common. 
Our rms estimates seem comparable to the 
scatter in 
\citet[][$\drms \approx 10-15\%$, compare to their Figure~3]
{KochanekDalal04}, 
though the average perturbation seems markedly different, 
even at a qualitative level.  Particularly, we find 
small ($|\avg{\dt}| \ll \drms$), negative average 
perturbations, whereas 
\citet[][]{KochanekDalal04} report large 
($|\avg{\dt}| \sim 10\% \sim \drms$), 
positive average perturbations.

How serious is the discrepancy between the model and 
observed lenses and what is it telling us?  
To address this question, we 
consider in some more detail our expression for 
the average perturbation.  We now show that 
{\em small negative perturbations are a 
generic prediction of linear perturbation
theory when the total mass of the 
lensing potential is held fixed.}  
Consider again our expression for the total magnification 
perturbation, Equation~\ref{eq:totpert}.  
For simplicity, assume that all perturbers are identical.  
Consider the shear contribution to the average perturbation.  
Two perturbers equidistant from an image and located at  
position angles $\phi$ and $\phi+\pi/2$ create equal and
opposite shears so the net shear contribution to the average
perturbations is zero, to a first approximation.  
The only nonzero contribution is caused by the convergence 
perturbations, so we can write
\begin{equation}
\label{eq:dtconv}
\avg{\dt} \simeq 2|\mu|(1-\kappa)\avg{\delta\kappa},
\end{equation}
where the expectation value of the convergence 
perturbation is yet to be determined.  
An argument similar to the one in 
\S~\ref{section_multiple_perturbers} results in 
\begin{equation}
\avg{\delta\kappa}= \sum_N P(N)N \int_{\cal{R}} 
	\frac{d^2\theta}{A(\cal{R})} \delta\kappa(\theta),
\label{eq:avg_delta_kappa_0}
\end{equation}
where the region $A( \cal{R} )$ is defined as before 
(i.e. $\btheta \in { \cal{R} }$ iff 
$\dmin < \delta( \btheta ) < \dnl$)
and $N$ is the number of perturbers.  
We assume that $P(N)$ is a Poisson distribution 
with an expectation value $\avg{N} = s A( \cal{R} )$, 
where $s$ is the mean number density of perturbers.  
Observe that if the integral over $\kappa$ in the 
above expression were extended over all space, 
we would recover the mass of the perturber.  
This leads us to define the nonlinear and linear mass 
fractions as
\begin{equation}
\label{eq:fnl}
\fnl = \frac{\Sc\dl^2}{m}\int_{|\delta|<\dnl} d^2\theta\ 
	\delta\kappa(\theta),
\end{equation}
and $\fl=1-\fnl$ respectively, where 
$m$ is the perturber mass.  Thus, $\fnl$ 
is the fraction of the perturber mass 
contained within the nonlinear region 
around an image for a perturber centered
on the image.  Using these definitions, 
and taking the limit $\delta_{min}\rightarrow 0$, 
Eq.~(\ref{eq:avg_delta_kappa_0}) becomes
\begin{equation}
\label{eq:meandk}
\avg{\delta\kappa} = \avg{N}\frac{m(1-\fnl)}{\Sc\dl^2A} 
	= \kappa_s (1-\fnl),
\end{equation}
where $\kappa_s = sm/(\Sc\dl^2)$ is just the 
convergence contribution by substructures of mass $m$.  
Replacing this contribution in our expression 
for the average perturbation, 
and including the negative mass
component $-\kappa_s$, we obtain
\begin{equation}
\label{eq:dtLIN}
\avg{\dt}_{\mathrm{LIN}} = -2 |\mu| (1-\kappa)\kappa_s \fnl.
\end{equation}
Thus, the average perturbation due to linear perturbers 
is always expected to be negative and approaches 
zero in the limit that local, linear perturbation 
theory holds.  Indeed, our choice $\lambda_a=4$ for 
Figure~\ref{moments} was made to make $f_{NL}$ small
so as to minimize nonlinear effects.
Lowering $\lambda_a$ increases  the nonlinear fraction
$f_{NL}$ and hence $\avg{\dt}$ becomes more {\em negative}
($f_{NL}\propto1/\sqrt{\lambda_a}$ at fixed mass).

It is important to emphasize that a negative 
$\avg{\dt}$ implies that positive-parity images 
will be dimmed and negative-parity images will be 
brightened, which is {\em opposite} 
from the trend for observed lenses 
\citep[][]{KochanekDalal04}, 
{\em provided} one uses the predicted macromodel 
magnification as a proxy for the observed magnification if 
the substructures were replaced by a smooth component.  
Because we assumed only that perturbations are linear and local,  
the implications are that either real lenses are significantly 
affected by nonlinear perturbers, perturbations are 
at least partly due to nonlocal perturbers, 
or the smooth macromodel 
magnification is a poor proxy for what the magnification 
would be in the absence of substructures.  Note that these 
various possibilities are {\em not} mutually exclusive. 
Of course, an alternative, and perhaps more contentious, 
possibility is simply that substructures within lens 
galaxies are not responsible for the observed 
magnification perturbations.
Assuming this last possibility is incorrect,
what is the most likely explanation for our results?

Nonlocal perturbations do not appear to be 
the most reasonable resolution.  
So long as linear perturbation 
theory holds, our na{\"{\i}}ve expectation 
is that the average convergence 
perturbations, and hence the average 
magnification perturbation itself, 
should be close to zero.  
Further, the two-dimensional surface number density 
of substructures predicted for CDM halos is a very 
slowly-varying function of radius out to several times 
the Einstein radius of the lens, so a constant perturber 
density seems like a reasonable approximation.  
Thus, we do not view nonlocal (non-constant substructure
surface density) perturbations as a viable explanation 
for the observed discrepancy.  However, it is worth noting 
that if nonlocal perturbations play a significant role, 
na{\"{\i}}ve comparisons of the substructure mass fractions 
derived from observational studies to those in 
simulations and semi-analytic models at radii 
of order a few $\kpc$, as have been performed so far 
\citep[e.g.,][]{zentner_bullock03,Maoetal04}, 
do not constitute fair comparisons.  
This alone warrants further study of the effects 
of nonlocal perturbers.

\begin{figure}[t]
\epsscale{1.0}
\plotone{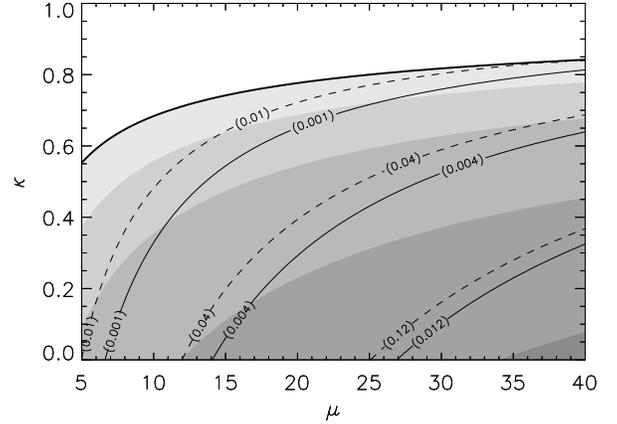}
\caption{
The expected number of substructures for various conditions.
The {\em unshaded} region above the thick, solid line is
unphysical for positive parity images.
The {\em shaded} regions below are obtained by computing the expected 
number of perturbers that create perturbations stronger than 
$\deltap$ (see Figure~\ref{y1}), and correspond, from top to bottom,
to $\avg{N|\delta>\delta_p}\in[0,20],[20,26],[26,29],[29,30],$ and
$30$ or more perturbers.  All substructures
parameters are as per section \ref{predictions}.
We loosely interpret  $\avg{N|\delta>\delta_p}$ as the
number of substructures found within 
a tidal radius of the image (see text).  
The {\em solid} contours are for 
$\avg{ N | \delta > 1}$, 
and give the expected number of nonlinear perturbers, which 
is seen to by quite small, $\lesssim 10^{-2}$.  The
{\em dashed}  contours are for a more conservative linear
cutoff $\dnl=20\%$, for which we find
$\avg{N|\delta>20\%}\lesssim 0.1$.
Truncated SIS profiles give results similar 
to that of the pseudo-Jaffe profiles.  
}
\label{num_subs}
\end{figure}

What about nonlinear perturbers?
In Figure~\ref{num_subs}, we show the expected number 
of non-linear perturbers $\avg{N|\delta>\dnl}$, 
for $\dnl=1$.  It is clear that this number is 
quite small, typically 
$\avg{N|\delta > 1} \lesssim 10^{-2}$.  
Even for a linear cutoff as small as 
$\dnl = 20\%$, the number of nonlinear perturbers 
is $\avg{N|\delta > 20\%} \lesssim 10^{-1}$.  The
prevalence of nonlinear perturbers is unlikely, 
unless observational biases preferentially select systems 
where nonlinear effects are large, in which case 
more sophisticated methods must be used to compare 
theoretical predictions to observational data, 
perhaps using Monte Carlo methods like those of 
\citet{zentner_bullock03} directly to model lens systems.  
While our results depend on our particular choice of 
profiles, it seems unlikely that they underestimate the 
number by significantly more than an order of magnitude, 
which is necessary to make nonlinear perturbers likely.  
Nevertheless, including many more massive substructures 
might add significantly to the number of nonlinear perturbers, 
as would adopting significantly higher substructure mass 
fractions, though higher mass fractions seem unlikely in 
the context of CDM.  

Even in such a case, it is difficult to imagine that the 
number of nonlinear perturbers could be greater than a few. 
We would then expect the contribution 
to the magnification perturbation by nonlinear
perturbers to have a large rms scatter so that 
$\drms \sim \avg{\dt}$.  
Because the scatter from linear 
perturbers appears to be broadly consistent 
with observed values, we expect it to be 
difficult to explain a nonzero net magnification
perturbation with nonlinear perturbers without 
at the same time increasing the scatter in the 
perturbations beyond that of observations. 

If neither nonlocality of the perturbations nor 
nonlinear perturbers can explain how 
positive- (negative-) parity images get 
brighter (dimmer) on average, then the 
only remaining possibility that is consistent with 
perturbations caused by CDM substructure 
is that for any given lens, 
the best-fit smooth model prediction for the 
magnification is a poor proxy for what the 
magnification would be if the mass in 
substructures were redistributed as a 
smooth component.  
We address these problems in detail in a 
forthcoming companion paper 
(Chen \& Rozo 2005, In preparation).

Before proceeding, we emphasize the importance
of tidal truncation of substructures in the expected number
of nonlinear perturbers.  In particular, if we assume pure
SIS perturber profiles,
the number of perturbers in the region $|\delta| > 20\%$ 
increases by a factor of $\sim 10$ and becomes 
of order $\sim 1$, as was demonstrated by 
\citet[]
[compare to our Figure~\ref{num_nonlin_ptmass_pert}]
{ChenKravtsovKeeton}.  
This reflects the fact that, in our model, 
$\sigma_{\mathrm{SIS}}( 20\% ) \gtrsim a^2$ 
for substructures of mass $\mmax$, 
implying that substructures become tidally 
truncated within the region 
$\delta_{SIS} > 20\%$.  Thus, 
use of simple SIS profiles that are not 
tidally-truncated may overestimate the importance 
of nonlinear substructures.  Moreover, this implies 
that in order to draw robust conclusions from lens 
systems, the effective truncation of substructures 
must be well understood in systems with a variety 
of different formation histories, a problem which is 
theoretically challenging.

%
%
%
\subsection{Individual Lenses}
\label{sub:ind}

Though the model that we presented is 
simplified and our results should generally be 
interpreted on a statistical basis, it is interesting 
to consider how this analysis can aid in the 
understanding of individual systems.  
For instance, if the probability of finding a 
substructure within a tidal radius of an image is small,
then the average perturbation ought to be a rather 
poor estimator of the most likely perturbation.  
In particular, the geometric weighting 
$\theta^2$ of substructures a distance $\theta$ 
away from the image is clearly not appropriate in 
this case.  Conversely, a large number of perturbers 
within a tidal radius would allow us to invoke the 
central limit theorem, in which case the probability 
distribution $\rho(\dt)$ would be well approximated 
by a gaussian distribution.  This raises a natural question.  
How many perturbers are found within a tidal radius or, 
more precisely, how many perturbers are found in the 
region $\delta > \deltap$?  In what follows, 
we refer to these two conditions interchangeably.  
The rigorous condition will always be 
$\delta > \deltap$, and we loosely interpret 
this as perturbers found 
within a tidal radius from the image.

Figure~\ref{num_subs} shows that for typical 
lens systems, the expected number of substructures with 
$\delta > \deltap$ is of order $\sim 20$  when
all substructure parameters are as described in 
\S~\ref{predictions}.  
The exact number is somewhat sensitive to the cutoff and
density profile.  For instance, at $\lambda_a=1$
the number of perturbers within a tidal radius decreases
to a few to several perturbers, and increases again to
$\sim 10$ perturbers if one lowers $\mmax$ by an order
of magnitude.  
It seems likely then that in most lens systems 
the total magnification perturbation is due 
to from a few to about twenty substructures 
contributing at approximately comparable levels.  
Thus we expect that actual perturbations may differ 
substantially from the average perturbation, 
implying a {\em large variance} 
(see Figure~\ref{moments}) 
and a {\em non-gaussian probability distribution} 
for the magnification perturbations. 

%
%
%
\section{Discussion and Conclusions}
\label{discussion}

The most surpising result from our analysis is 
that at a {\em fixed lens mass}, 
introducing a clumpy component tends 
to demagnify positive-parity images 
while magnifying negative-parity images on average, 
with the net effect tending towards 
zero as nonlinear perturbers become less important.  
This does not imply that the traditional arguments 
that {\em adding} substructure will, on average, 
increase the magnification of positive-parity images 
are wrong.  Rather, it emphasizes that the increase in 
magnification of the observed image is, 
in part, due to the simple fact that mass is being 
added to the lens.

If one adopts the predicted magnification from the 
best-fit smooth macromodel to a lens as a proxy for what 
the magnification would have been if the substructure were 
not present and the lens were smooth, 
the observed average perturbation is positive, 
{\em opposite} from what we predict.  
For low substructure mass fractions (comparable to the 
level predicted by the CDM paradigm of structure formation), 
nonlinear perturbers are unlikely and 
hence nonlinearities, which were disregarded in our study, 
are not a likely explanation for the observed discrepancy.  
Further, if the scatter due to linear perturbers is 
found to broadly match that in observations, 
it will be difficult for nonlinear perturbers 
to explain the observed average perturbation 
without, at the same time, significantly 
increasing the scatter.  
This suggests that the use of the best-fit smooth model 
magnification as a proxy for the magnification of a 
smooth lens may not be justified.
We investigate this specific question in more 
detail in a follow-up study 
(Chen \& Rozo 2005, in preparation).

Another surprising aspect of our analysis is the 
possibility of having sizable magnification 
perturbations arising from small, but coherent, 
contributions from several substructures within 
the lens halo.  Such perturbations are difficult 
to account for in individual lenses because 
knowledge of the expected non-gaussian distribution 
$\rho(\dt)$ of the perturbations is necessary to 
perform maximum likelihood analyses of particular 
systems 
\citep[e.g.,][]{MaoSchneider98}.
Likewise, the conclusions of analyses that assume a 
single substructure is responsible for the full 
magnification perturbation 
\citep[e.g.,][]{Keeton01,DoblerKeeton05} 
can change when multiple perturbers are present.
As a simple example, the relation between 
magnification perturbation cross section and 
the probability of an image being perturbed is 
weakened if large perturbations can arise from 
the coherent contribution of multiple, weak perturbations.

We emphasize that the importance of coherent 
perturbations depends on the tidal radii of 
substructures.  This result underlines the 
importance of numerical studies such as 
those by \citet{Bradacetal04} and 
\citet{Amaraetal04}, where tidal limitation of 
substructures is inherent in the calculation.   
Yet the theoretical challenges remain 
significant.  The fact that we expect low-mass 
substructures (i.e. substructures with masses below 
the resolution limit of present day numerical 
simulations) to contribute to magnification 
perturbations at a non-negligible level and the 
fact that the scatter in the properties of 
substructure from halo to halo is significant 
both argue for complementing direct numerical 
simulation with analytic \citep[e.g.,][]{chen_etal03} 
and semi-analytic computations \citep[e.g.,][]
{zentner_bullock03,zentner_etal05c,taylor_babul05}.  
Of course, the drawback of these studies is that 
substructure dynamics and tidal truncation are 
handled only in an approximate way.

One issue that was brought up in our analysis was 
the definition of the substructure mass fraction.  
In particular, we focused on magnification perturbations 
induced by granularity, so we argued that for a host halo 
with with a mass $M_E$ within its Einstein radius,
one is interested in the mass fraction of substructures 
with mass $m \lsim \mmax$, where 
$\mmax \approx 10^{-2} M_E \sim 10^{-4} \Mvir$.  
Unfortunately, the huge dynamic range that is necessary 
to probe the properties of halo substructure with 
$N$-body simulations limits their ability 
to probe substructures to masses
{\em larger} than $m \gtrsim 10^{-5} \Mvir$ 
\citep[e.g.,][]{Maoetal04} and to limit such 
studies to a relatively small number of systems.  
Therefore the CDM predictions for 
the appropriate substructure values 
require some extrapolation and are thus may be 
robust than one might hope.  
There are other theoretical uncertainties as well.  
In particular, \citet{Oguri04} pointed out that in 
systems where the lens galaxy is a member of a group 
or a cluster, the relevant substructure mass fraction
should include contributions from substructures within 
the group halo itself.  
Likewise, anisotropy of the host halo and 
its substructure population 
\citep[e.g.,][]{zentner_etal05a,kang_etal05,libeskind_etal05} 
leads to significant systematic variations in substructure 
mass fractions along various lines of sight and 
may alter the expectation value of the substructure 
mass fraction due to projection effects 
(Zentner, Rozo, \& Kravtsov 2005, In preparation).  
Finally, the abundance of subhalos in CDM cosmologies 
depends on the amplitude of the power spectrum on 
small scales \citep{zentner_bullock03}, and may depend 
as well on baryonic processes within the host 
halo that are yet to be fully understood.  
More theoretical work is needed to fully characterize 
the CDM predictions for the substructure mass fractions 
relevant for lensing.

While the detection of dark substructures through 
image splitting or magnification perturbations of 
extended sources 
\citep{InoueChiba03}, would 
provide fairly unambiguous evidence of CDM substructures, 
until such detections become feasible our 
only way to probe these dark substructures is through their
effects on multiply-imaged sources.  
This makes unambiguous detection of dark substructures 
in individual lenses more difficult because more ``mundane'' 
causes for the small-scale perturbations to the 
lensing potentials exist.  
Spectroscopic analyses of lenses may go 
some way in removing these ambiguities
\citep{MoustakasMetcalf03},
especially if complemented by studies of 
the associated astrometric perturbations 
(Chen et al. 2005, in preparation) and by 
looking for distortions in lensed jets 
\citep[][]{MetcalfMadau01}.
In light of these ambiguities, we believe that 
if CDM were to provide concrete predictions for 
the statistical properties of the magnification 
perturbations induced by substructures, 
confirmation of these properties in observed lens 
samples would provide strong evidence in support 
of the CDM paradigm.  In this paper, we have identified 
one way (granularity) in which CDM substructures can 
produce flux perturbations, and shown analytically that 
a proper understanding of the substructure profiles, 
including their tidal limitation, and the mass spectrum 
of subhalos are all necessary components of a faithful 
treatment of the magnification perturbations and to 
obtain accurate estimates of substructure abundance.
Nevertheless, we also expect other effects such as 
perturbations by massive substructures 
\citep{CohnKochanek04} to play important roles, 
though the corresponding statistical properties of 
these perturbations remain to be understood.  While 
a complete of understanding of the CDM predictions 
for the statistical properties of magnification perturbations 
is a difficult task, it is one which is likely to
be of vital importance for testing 
the CDM paradigm at small scales and 
in the highly-nonlinear regime.

\acknowledgments

We would like to thank Scott Dodelson for useful 
discussions throughout the development of this work, 
and Charles Keeton for a carfeul reading and critique 
of this manuscript.  We also thank Andrey Kravtsov 
for valuable comments on the manuscript.  
ER, ARZ, and JC are supported by 
The Kavli Institute for Cosmological Physics 
at The University of Chicago and by the National 
Science Foundation under grant NSF PHY 0114422. 
GB is supported by the DOE and the NASA grant NAG 5-10842 
at Fermilab.  This research made use of the NASA 
Astrophysics Data System.

%
%
\appendix

%
%
\section{A. When Can Astrometric Perturbations Be Neglected?}
\label{appendixa}

In general, if the potential of a macrolens is perturbed, 
the image position of a source will change in accordance with 
the perturbed lens equation.  We wish to 
determine under what conditions is it possible 
to neglect such perturbations in comparison to the 
magnification perturbations.  

Consider a lens system where the unperturbed image position 
of a point source is $\btheta$.  Neglecting astrometric 
perturbations, the only change to the convergence and shear 
sampled by the image are the values 
$\delta\kappa$ and $\delta\gamma$ of 
the perturber at the image position.  
Relative to this first-order perturbation, 
there are higher order corrections due to the 
image being displaced because both
the macromodel and perturber convergence and 
shear at the new image position will be different.  
For these higher order terms to be negligible, 
it is required that 
\begin{equation}
\delta\kappa \gg 
\frac{\partial (\delta \kappa)}{ \partial \btheta} \dbtheta 
\hspace{0.1 in} \mbox{ and } \hspace{0.1 in}
\delta\gamma\gg 
\frac{ \partial \gamma}{ \partial \btheta} \dbtheta,
\label{cond}
\end{equation}
where $\dbtheta$ is the position perturbation, 
and similar inequalities must hold for the image shear.  
To determine when these conditions are met, 
one needs to compute the perturbation $\dbtheta$. 

Let a spherically-symmetric perturber be introduced at 
a position $\bthetap$.  
The perturbed lens equation is 
\begin{equation}
\label{eq:pertlens}
\btheta + \dbtheta = \bbeta +  \balpha(\btheta+\dbtheta) 
	 + \dbalpha(\btheta+\dbtheta).
\end{equation}
$\dbalpha$ is the deflection due to the perturber, which is 
a function of the relative position of the perturber 
$\btheta+\dbtheta-\bthetap$.  
Linearizing $\balpha$ and using the unperturbed lens 
equation $\btheta=\bbeta+\balpha(\btheta)$, we obtain
\begin{equation}
M^{-1}\dbtheta=\dbalpha(\btheta+\dbtheta),
\label{pospert}
\end{equation}
where $M^{-1}$ is the inverse magnification tensor due 
to the macromodel, 
$M^{-1}_{\phantom{-1}ij} = 
\delta_{ij} - \partial \alpha_{i}/\partial \theta_{j}$.  
To get this equation we only needed to assume that the 
macrolens properties vary slowly over a distance 
$|\delta\theta|$.  To fully linearize the system and set 
$\dbalpha(\btheta+\dbtheta) \approx \dbalpha(\btheta)$,
the system must satisfy the more stringent constraint
\begin{equation}
|\dbalpha| \gg \Big| 
\frac{\partial\dbalpha}{\partial\btheta} \dbtheta\Big|.
\label{lincond}
\end{equation}
Assuming we can fully linearize the system and 
inverting we obtain 
\begin{equation}
\label{eq:dthetaM}
\dbtheta=M\dbalpha.
\end{equation}
However, there is a self-consistency constraint 
provided by Equation~\ref{lincond} that must be satisfied.  
Inserting the fully-linearized solution into 
Equation~\ref{lincond}, we find that one must have 
\begin{equation}
|\dbalpha|\gg 
\Big| \frac{\partial\dbalpha}{\partial\btheta}M\dbalpha\Big|.
\label{lincond1}
\end{equation}

Consider the matrix 
\begin{equation}
\frac{\partial\dbalpha}{\partial\btheta}M =
	\left( \begin{array}{cc}
		\delta\kappa+\delta\gamma_1	& \delta\gamma_2 \\
		\delta\gamma_2			& \delta\kappa-\delta\gamma_1 
	\end{array} \right)
	\left( \begin{array}{cc}
		1/\lambda_T 	& 0 \\
		0		& 1/\lambda_R
	\end{array} \right),
\end{equation}
where $\lambda_{R} = 1 - \kappa + \gamma $ and 
$\lambda_T=1-\kappa-\gamma$.
Typically, one is interested in images that form
near the tangential critical curve of the macrolens, 
which satisfy $1/\lambda_T \gg 1/\lambda_R$.  
Setting $1/\lambda_R = 0$ in the above
expression leads to 
\begin{equation}
\frac{\partial\dbalpha}{\partial\btheta}M \approx 1/\lambda_T
	\left( \begin{array}{cc}
		\delta\kappa+\delta\gamma_1	& 0 \\
		\delta\gamma_2			& 0
	\end{array} \right).
\end{equation}
The eigenvalues of this matrix are 
zero and $(\delta\kappa+\delta\gamma_1)/\lambda_T$.
Letting $\eunit$ be the unit eigenvector 
corresponding to the nonzero eigenvalue,
it follows that
\begin{equation}
\Big|\frac{\partial\dbalpha}{\partial\btheta}M\dbalpha\Big|
	\approx \frac{1}{\lambda_T}(\delta\kappa+\delta\gamma_1) 
		(\eunit \cdot \dbalpha)
	<  \frac{1}{\lambda_T}(\delta\kappa+\delta\gamma)|\dbalpha|.
\end{equation}
Inserting this result back into Equation~(\ref{lincond1}), 
we find that full linearization of the problem is 
self consistent when
\begin{equation}
\frac{1}{\lambda_T}(\delta\kappa + \delta\gamma) \ll 1
\label{lincond2}
\end{equation}

Now that we have found a consistent way 
to estimate the astrometric perturbation to the 
image position, we can estimate the 
change in macromodel and perturber convergences 
between the original, unperturbed image position 
and the new image position.  

Consider first the macromodel.  
If $\theta_E$ is the Einstein radius of the macromodel, 
one typically expects 
$\partial \kappa / \partial \theta \approx \kappa / \theta_E$ 
for the macromodel, and 
$\delta \alpha \approx \delta \kappa \theta$ 
for the deflection angle due to the perturber.  
Inserting these into Eq.~(\ref{cond}) and using 
$\delta \theta \approx M \delta\alpha \approx 
\lambda_T \delta\kappa\theta$, 
we find that position perturbations are 
negligible as far as the macromodel is concerned 
provided
\begin{equation}
\theta \ll \lambda_T \theta_E \lesssim 0.1\theta_E.
\end{equation}
That is, position perturbations due to nearby perturbers 
are expected to be negligible 
(relative to the convergence perturbation $\delta\kappa$).
The corresponding condition for the 
perturber is 
\begin{equation}
\frac{\delta\kappa}{\lambda_T} \ll 1,
\end{equation}
a constraint that is already contained within the 
self-consistency requirement given by 
Eq.~(\ref{lincond2}).  A similar computation for 
the shear yields the exact same conclusions.
It follows that if the linearization condition, 
Eq.~(\ref{lincond2}), holds, we may neglect 
astrometric perturbations when computing 
magnification perturbations.

We emphasize that if an image lies very close 
to the tangential critical curve, the
condition expressed in Eq.~(\ref{lincond2}) 
may represent a very stringent requirement.  
For order of magnitude purposes, one can assume
$\delta\kappa \sim \delta\gamma$ and $\lambda_R \sim 1$, 
so that our requirement for neglecting astrometric 
perturbations can be roughly expressed as
\begin{equation}
\delta\kappa+\delta\gamma \ll 1/|\mu|.
\label{lincond3}
\end{equation}
In particular, this means that the maximum value that a 
perturbation $\delta\kappa$, or $\delta\gamma$, 
can take while still producing negligible 
astrometric perturbations is of order $|\mu|^{-1}$.  
It is possible to go even further 
and relate this to flux perturbations.  
We saw in Eq.~(\ref{eq:smallfluxpert}) 
that for small perturbations $\delta\kappa$ 
and $\delta\gamma$,
the flux perturbation is given by
\begin{equation}
\frac{\delta\mu}{|\mu|}=2|\mu|\big\{ (1-\kappa)\delta\kappa
	+\gamma\delta\gamma_1 \big\}.
\end{equation}
One expects in general that $\kappa \sim \gamma \sim 1/2$, 
so that the relative flux perturbation 
$\delta\mu/|\mu|$ is of order
\begin{equation}
\label{eq:mucond}
\frac{\delta\mu}{|\mu|}\sim |\mu|(\delta\kappa+\delta\gamma).
\end{equation}
The linearization condition we derived, 
Equation~(\ref{lincond3}), states that the 
right-hand side of Eq.~(\ref{eq:mucond}) ought 
to be much smaller than one. 
Hence, another way of phrasing the 
linearization condition is simply
\begin{equation}
\frac{\delta\mu}{|\mu|} \ll 1.
\end{equation}
That is, astrometric perturbations to the image 
position may be neglected when considering
small flux perturbations.

%
%
\section{B. Point Mass Perturbers}
\label{ptmass}

Consider the perturbations due to a point mass of mass $m$.  
The perturbing potential, convergence, and shear are 
\begin{equation}
\delta\psi(\theta) = b^2 \ln (\theta), 
\hspace{0.1 in} \delta\kappa(\theta) = 0, \hspace{0.1 in} 
\mbox{and} \hspace{0.1 in} \delta\gamma(\theta) 
= \frac{b^2}{\theta^2}, 
\label{ptmass-profiles}
\end{equation}
where $b^2=m/\pi\Sc\dl^2$ is the 
Einstein radius of the point mass.
Inserting these expressions in 
Equations~(\ref{eq:Theta}), we find only
one root so that 
\begin{equation}
\tmax^2(\Delta) = \frac{2|\mu|\gamma}{|\Delta|}b^2 \hspace{0.5 in}
	\mbox{and} \hspace{0.5 in} \tmin^2(\Delta)=0
\label{ptmassTheta}
\end{equation}
for both $\Delta>0$ and $\Delta<0$.  Likewise, 
inserting the shear and convergence profiles in 
Equation~\ref{eq:phi_m}, we find 
\begin{equation}
2\phim(\Delta|\theta) = \cos^{-1}
	\left( -\frac{\Delta}{|\Delta|}
        \frac{\theta^2}{\tmax^2}\right).
\end{equation}
%
Using the above expressions and carrying out the integrals in 
Equation~(\ref{eq:cross_section}), we find
\begin{equation}
\sigma(\Delta) = \frac{2\gamma|\mu|}{|\Delta|}b^2.
\label{ptmass_cs}
\end{equation}

This holds for both $\Delta>0$ and $\Delta<0$; the cross sections 
for brightening and dimming an image are exactly equal to each other.  
This is to be expected: point mass perturbers couple only through
the shear, so perturbers placed at angles $\phi$ and 
$\phi + \pi/2$ create equal and opposite perturbations.  
Since no such symmetry exists for perturbers with 
nonzero convergence, the equality 
$\sigma(\Delta) = \sigma(-\Delta)$ 
does not hold in general.  
This symmetry also implies that $y_1(\delta) = 0$ 
and hence the average magnification perturbation due 
to point masses is zero
\footnote{
This is {\em not} correct when nonlinear perturbations are likely, 
in which case negative-parity images have a large probability of 
being strongly demagnified.  
Positive-parity images are also demagnified on average, 
but the effect is not as strong as for negative-parity images 
\citep[][]{SchechterWambsganss02}.}.
However, the variance is nonzero.  
Inserting Eq.~(\ref{ptmass_cs}) 
into Eq.~(\ref{eq:vardt}), 
and using the definition of the 
Einstein radius $\pi\Sc\dl^2 b^2 = m$, 
we obtain
\begin{eqnarray}
\Var{\dt} & = & \int dm \frac{1}{\Sc}\frac{ds}{dm}m 
\frac{4}{\pi}\gamma|\mu|\Delta \\
	& = & \frac{4}{\pi}\gamma\kappa|\mu|\Delta \fsat = 
\avg{N|\Delta}\Delta^2, 
\end{eqnarray}
where we have defined the perturber mass fraction $\fsat$, 
such that $\fsat \kappa \Sc = \int dm \frac{ds}{dm}m$ and 
$\avg{N|\Delta}$ is the expected number of perturbers 
that produce a perturbation stronger than $\Delta$.  
Interestingly, the variance of the perturbation does not 
depend in any way on the perturber mass spectrum.


\begin{figure}[t]
\epsscale{1.0}
\plotone{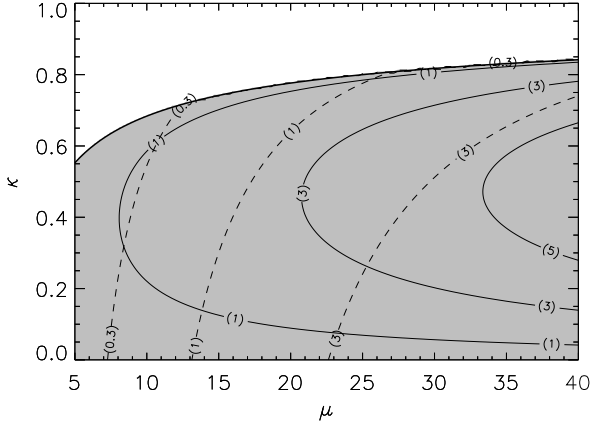}
\caption{
The {\em solid} contours give the expected number of stars in a 
lensing galaxy which produce strong (nonlinear) 
magnification perturbations on an image.  A linear 
cutoff $\dnl = 0.2$ and a stellar mass fraction $f_* = 0.1$ 
at the galaxy's Einstein radius are assumed.  
The {\em lightly-shaded} area represents the allowed region of 
parameter space for positive-parity images, 
and contours are for $1,3,$ and $5$ nonlinear stars.  
We conclude that stellar magnification perturbations 
to multiply-imaged quasars 
will generally be large and nonlinear 
(hence microlensing).  The {\em dashed} contours
illustrate the number of mildly nonlinear 
($\delta > 20\%$) SIS perturbers predicted 
by an SIS model assuming that the number 
density of perturbers is equal to that 
assumed in \S~\ref{predictions} for pseudo-Jaffe profiles.  
This is to be compared to the dashed lines in 
Fig.~\ref{num_subs}.  The number of nonlinear SIS 
perturbers is of order unity, whereas the number of 
nonlinear perturbers for the truncated profiles is 
$ \lesssim 0.1$.  This reflects the fact that 
tidal truncation occurs within the region 
$\delta>20\%$ for an SIS perturber.  
We conclude that SIS profiles lead to 
strong overestimates of the importance of 
nonlinear perturbers.
}
\label{num_nonlin_ptmass_pert}
\end{figure}


\it Lensing by stars: \rm 
As a particular application, consider 
magnification perturbations due to stars 
in the lens galaxy.  We take all stars to be 
identical in mass and define $f_*$ as the surface 
mass fraction of stars at the image position, 
$f_* = \Sigma_*/\Sigma_{Tot}$.  
The expected number of stars in the region 
$|\delta| >\Delta$ (note the absolute value) 
is thus
\begin{equation}
\avg{N||\delta|>\Delta} = \frac{f_* \kappa \sc}{m_*}\dl^2 
		\left\{\sigma(\Delta)+\sigma(-\Delta)\right\}
		= \frac{4}{\pi}\frac{|\mu|\gamma\kappa}{\Delta} f_*
\label{num_star_pert}
\end{equation}

For $f_* \gtrsim 0.1$, 
$\kappa \approx \gamma\approx 1/2$, and 
$\mu\approx 10$, 
we get $\avg{N||\delta|>1} \gtrsim 0.3$.  
We expect then, that nonlinear perturbers will play 
an important role in determining the observed 
magnification of an image, a result
that has been amply demonstrated in previous studies 
\citep[see for example][]{SchechterWambsganss02}


\section{C. Singular Isothermal Sphere Perturbers}
\label{sis}

Consider perturbations by a singular isothermal sphere.  
The perturbing potential, convergence, and shear profiles are
\begin{equation}
\delta\psi(\theta) = b \theta, \hspace{0.1 in}
\delta\kappa = \frac{b}{2\theta}, \hspace{0.1 in} \mbox{and} \hspace{0.1 in}
	\delta\gamma = \frac{b}{2\theta},
\label{sis-profiles}
\end{equation}
where $b$ is the Einstein radius of the perturber, 
given by
\begin{equation}
b = 4\pi\left(\frac{\sigma_v}{c}\right)^2 \frac{\dls}{\ds},
\end{equation}
and $\sigma_v$ is the velocity dispersion of the perturber.  
From Equation~(\ref{eq:Theta}), 
the nonzero solutions for $\tmax$ are
\begin{equation}
\tmax(\Delta) =  \frac{|\mu|b}{|\Delta|}\times \left\{ \begin{array}{lc}
	\lambda_R \mspaces & \mbox{if}\ \Delta>0\ \mand\ \lambda_R\ge 0 \\
	-\lambda_T \mspaces & \mbox{if}\ \Delta<0\ \mand\ \lambda_T \le 0. \\
		\end{array} \right .
\end{equation}
$\tmax=0$ for $\Delta>0,\lambda_+<0$ and for 
$\Delta<0,\lambda_->0$, and $\tmin=0$ always.
Note $\tmax(\Delta)=0$ implies that perturbations 
of size $\Delta$ are never realized, so 
it is impossible for SIS perturbers to either dim a 
positive-parity image or brighten a 
doubly negative-parity image.  Finally, from 
Equation~(\ref{eq:phi_m}), we find
\begin{equation}
2\phi_m(\theta|\Delta) = \cos^{-1}\left(
		\frac{1-\kappa}{\gamma} - 
              \frac{1}{\gamma|\mu|} \frac{\theta}{b}\Delta \right).
\end{equation}

Inserting our above expressions into 
Equation~(\ref{eq:cross_section}), and 
defining the amplitude $\sigma_0$ via 
\begin{equation}
\sigma_0=\frac{\pi}{2}\frac{|\mu|^2b^2}{\Delta^2},
\end{equation}
we obtain
\begin{itemize}
\item $\Delta>0,\lambda_->0,\lambda_+>0:\ $ Brightening of a positive-parity image.
	$$\sigma=\sigma_0\left[2(1-\kappa)^2+\gamma^2\right]$$
\item $\Delta>0,\lambda_-<0,\lambda_+>0:\ $ Dimming of a negative-parity 
	image.\footnote{Even though
	the result we quote here looks quite different from the one 
        presented in Equation~17 of \citet{Keeton03}, it is simple to 
        show that 
        $\pi-2\theta_\delta = \cos^{-1}\left(-\frac{|1-\kappa|}{\gamma}\right)$,  
	so that both expressions for $\sigma(\Delta)$ do indeed agree 
        to leading order in $\delta$ ($\theta_\delta$ is defined in \citet[][]{Keeton03}). 
	A similar argument holds for the case $\Delta<0,\lambda_T<0,\lambda_R>0$.}
	$$\sigma=\sigma_0\left\{ 
		\frac{1}{\pi}\cos^{-1}\left(-\frac{|1-\kappa|}{\gamma}\right) 
			\big[2(1-\kappa)^2+\gamma^2\big]+
		\frac{3}{\pi}\frac{1-\kappa}{|\mu|^{1/2}}\right\}$$ 
\item $\Delta<0,\lambda_-<0, \lambda_+>0:\ $ Brightening of a negative-parity image.
	$$\sigma=\sigma_0\left\{ 
		\frac{1}{\pi}\cos^{-1}\left(\frac{|1-\kappa|}{\gamma}\right) 
			\big[2(1-\kappa)^2+\gamma^2\big]-
		\frac{3}{\pi}\frac{1-\kappa}{|\mu|^{1/2}}\right\}$$.
\item $\Delta<0,\lambda_-<0,\lambda_+<0:\ $ Dimming of a doubly negative-parity image.
	$$\sigma=\sigma_0\left[2(1-\kappa)^2+\gamma^2\right]$$
\end{itemize}

The expressions above are in exact agreement with the results from 
\citet[][]{Keeton03} in the limit 
$|\Delta| \ll 1$.  In particular, the cross sections 
in \citet[][]{Keeton03} contain terms that are
higher order in $\Delta$, not found in our expressions.
However, as we already noted, these corrections are
important only when astrometric perturbations become 
non-negligible.  Since negligible astrometric perturbations 
were also assumed in \citet[][]{Keeton03}, these
 higher-order terms are important only when both 
formalisms break down.

Before moving on, it is worth noticing that while the 
cross sections we have derived depend on the parity of the 
image, this dependence is something of happenstance.
In particular, what differentiates the various 
expressions for the magnification perturbation 
cross sections is not the image parity, 
but whether Equation~(\ref{eq:Theta}) has zero, one, 
or two roots.  For an SIS perturber, these conditions 
coincide with the parity of the image, 
but this is not true in general.
Indeed, we already showed in Appendix~\ref{ptmass}, 
that point masses do not distinguish between positive- 
and negative-parity images.

Consider an ensemble of SIS perturbers, which we choose 
to parameterize through their Einstein radii $b$.  
Assuming the macroimage has positive parity, 
Equation~(\ref{eq:avgdt}) becomes
\begin{equation}
\avg{\dt} = \pi |\mu|^2[2(1-\kappa)^2+\gamma^2] 
\int db\ \frac{ds}{db} b^2
	\left[1/\delta_m-1/\Delta\right],
\end{equation}
which clearly diverges in the limit $\delta_m\rightarrow 0$.  
The physical reason behind this is clear: 
each individual perturber within a distance $\theta$ from
the image contributes a positive perturbation 
$\delta \sim 1/\theta$.
Since the number of perturbers in such a ring 
scales as $\theta$, the net perturbation is divergent.

How generic are these divergences?  
If the magnification perturbation is dominated by 
the convergence perturbation $\delta \kappa$, 
then perturbers along a ring will give 
nonzero average perturbations that scale as 
$\theta \delta \kappa$.  Convergence is
expected then if, and only if, $\delta \kappa$ 
falls faster than $1/\theta^2$.
Since any such perturber has a finite mass, 
we conclude that finite magnification perturbations 
are expected if and only if the perturbers have finite 
masses.

%
\bibliography{ms}

\begin{thebibliography}{103}
\expandafter\ifx\csname natexlab\endcsname\relax\def\natexlab#1{#1}\fi

\bibitem[{{Amara} {et~al.}(2004){Amara}, {Metcalf}, {Cox}, \&
  {Ostriker}}]{Amaraetal04}
{Amara}, A., {Metcalf}, R.~B., {Cox}, T.~J., \& {Ostriker}, J.~P. 2004,
  astro-ph/0411587

\bibitem[{{Baltz} {et~al.}(2000){Baltz}, {Briot}, {Salati}, {Taillet}, \&
  {Silk}}]{baltz_etal00}
{Baltz}, E.~A., {Briot}, C., {Salati}, P., {Taillet}, R., \& {Silk}, J. 2000,
  PRD, 61, 023514

\bibitem[{{Benson} {et~al.}(2002){Benson}, {Frenk}, {Lacey}, {Baugh}, \&
  {Cole}}]{benson_etal02}
{Benson}, A.~J., {Frenk}, C.~S., {Lacey}, C.~G., {Baugh}, C.~M., \& {Cole}, S.
  2002, MNRAS, 333, 177

\bibitem[{{Berezinsky} {et~al.}(1997){Berezinsky}, {Bottino}, \&
  {Mignola}}]{berezinsky_etal97}
{Berezinsky}, V., {Bottino}, A., \& {Mignola}, G. 1997, Physics Letters B, 391,
  355

\bibitem[{{Bergstr{\"{o}}m} {et~al.}(1999){Bergstr{\"{o}}m}, {Edsj{\"{o}}},
  {Gondolo}, \& {Ullio}}]{bergstrom_etal99}
{Bergstr{\"{o}}m}, L., {Edsj{\"{o}}}, J., {Gondolo}, P., \& {Ullio}, P. 1999,
  PRD, 59, 043506

\bibitem[{Bertone {et~al.}(2005)Bertone, Hooper, \& Silk}]{Bertone_etal05}
Bertone, G., Hooper, D., \& Silk, J. 2005, Phys. Rept., 405, 279

\bibitem[{{Blumenthal} {et~al.}(1984){Blumenthal}, {Faber}, {Primack}, \&
  {Rees}}]{blumenthal_etal84}
{Blumenthal}, G.~R., {Faber}, S.~M., {Primack}, J.~R., \& {Rees}, M.~J. 1984,
  Nature, 311, 517

\bibitem[{{Bode} {et~al.}(2001){Bode}, {Ostriker}, \& {Turok}}]{bode_etal01}
{Bode}, P., {Ostriker}, J.~P., \& {Turok}, N. 2001, ApJ, 379, 440

\bibitem[{{Bottino} {et~al.}(1998){Bottino}, {Donato}, {Fornengo}, \&
  {Salati}}]{bottino_etal98}
{Bottino}, A., {Donato}, F., {Fornengo}, N., \& {Salati}, P. 1998, PRD, 58,
  123503

\bibitem[{{Brad{\v{a}}c}(2002)}]{Bradacetal02}
{Brad{\v{a}}c}, M. 2002, A \& A, 388, 373

\bibitem[{{Brad{\v{a}}c} {et~al.}(2004){Brad{\v{a}}c}, {Schneider}, {Lombardi},
  {Steinmetz}, {Koopmans}, \& {Navarro}}]{Bradacetal04}
{Brad{\v{a}}c}, M., {Schneider}, P., {Lombardi}, M., {Steinmetz}, M.,
  {Koopmans}, L.~V.~E., \& {Navarro}, J.~F. 2004, A \& A, 423, 797

\bibitem[{{Bullock} {et~al.}(2000){Bullock}, {Kravtsov}, \&
  {Weinberg}}]{bullock_etal00}
{Bullock}, J.~S., {Kravtsov}, A.~V., \& {Weinberg}, D.~H. 2000, \apj, 539, 517

\bibitem[{{Calc{\'{a}}neo-Rold{\'{a}}n} \&
  {Moore}(2000)}]{calcaneo-roldan_moore00}
{Calc{\'{a}}neo-Rold{\'{a}}n}, C. \& {Moore}, B. 2000, PRD, 62, 123005

\bibitem[{{Chen} {et~al.}(2003{\natexlab{a}}){Chen}, {Jing}, \&
  {Yoshikaw}}]{chen_etal03}
{Chen}, D.~N., {Jing}, Y.~P., \& {Yoshikaw}, K. 2003{\natexlab{a}}, \apj, 597,
  35

\bibitem[{{Chen} {et~al.}(2003{\natexlab{b}}){Chen}, {Kravtsov}, \&
  {Keeton}}]{ChenKravtsovKeeton}
{Chen}, J., {Kravtsov}, A.~V., \& {Keeton}, C.~R. 2003{\natexlab{b}}, ApJ, 592,
  24

\bibitem[{{Chiba}(2002)}]{Chiba02}
{Chiba}, M. 2002, ApJ, 565, 17

\bibitem[{{Cohn} \& {Kochanek}(2004)}]{CohnKochanek04}
{Cohn}, J.~D. \& {Kochanek}, C.~S. 2004, ApJ, 608, 25

\bibitem[{{Cole} {et~al.}(1994){Cole}, {Aragon-Salamanca}, {Frenk}, {Navarros},
  \& {Zepf}}]{cole_etal94}
{Cole}, S., {Aragon-Salamanca}, A., {Frenk}, C.~S., {Navarros}, J.~F., \&
  {Zepf}, S.~E. 1994, MNRAS, 271, 781

\bibitem[{{Col{\'{\i}}n} {et~al.}(2000){Col{\'{\i}}n}, {Avila-Reese}, \&
  {Valenzuela}}]{colin_etal00}
{Col{\'{\i}}n}, P., {Avila-Reese}, V., \& {Valenzuela}, O. 2000, ApJ, 542, 622

\bibitem[{{Dalal} \& {Kochanek}(2002a)}]{DalalKochanek02a}
{Dalal}, N. \& {Kochanek}, C.~S. 2002a, Apj, 572, 52

\bibitem[{{Dalal} \& {Kochanek}(2002b)}]{DalalKochanek02b}
---. 2002b, astro-ph/0202229

\bibitem[{{De Lucia} {et~al.}(2004){De Lucia}, {Kauffmann}, {Springel},
  {White}, {Lanzoni}, {Stoehr}, {Tormen}, \& {Yoshida}}]{delucia_etal04}
{De Lucia}, G., {Kauffmann}, G., {Springel}, V., {White}, S.~D.~M., {Lanzoni},
  B., {Stoehr}, F., {Tormen}, G., \& {Yoshida}, N. 2004, \mnras, 348, 333

\bibitem[{{Dekel} \& {Silk}(1986)}]{dekel_silk86}
{Dekel}, A. \& {Silk}, J. 1986, ApJ, 303, 38

\bibitem[{{Diemand} {et~al.}(2004){Diemand}, {Moore}, \&
  {Stadel}}]{diemand_etal04}
{Diemand}, J., {Moore}, B., \& {Stadel}, J. 2004, \mnras, 352, 535

\bibitem[{{Dobler} \& {Keeton}(2005)}]{DoblerKeeton05}
{Dobler}, G. \& {Keeton}, C.~R. 2005, ApJ, Submitted (astro-ph/0502436)

\bibitem[{Donato {et~al.}(2004)Donato, Fornengo, Maurin, \&
  Salati}]{donato_etal03}
Donato, F., Fornengo, N., Maurin, D., \& Salati, P. 2004, Phys. Rev., D69,
  063501

\bibitem[{{Efstathiou}(1992)}]{efstathiou92}
{Efstathiou}, G. 1992, MNRAS, 256, 43P

\bibitem[{{Evans} \& {Witt}(2003)}]{EvansWitt03}
{Evans}, N.~W. \& {Witt}, H.~J. 2003, MNRAS, 245, 1351

\bibitem[{{Gao} {et~al.}(2004){Gao}, {White}, {Jenkins}, {Stoehr}, \&
  {Springel}}]{gao_etal04a}
{Gao}, L., {White}, S.~D.~M., {Jenkins}, A., {Stoehr}, F., \& {Springel}, V.
  2004, \mnras, accepted

\bibitem[{{Gaudi} \& {Petters}(2002)}]{GaudiPetters02b}
{Gaudi}, B.~S. \& {Petters}, A.~O. 2002, ApJ, 580, 468

\bibitem[{{Ghigna} {et~al.}(2000){Ghigna}, {Moore}, {Governato}, {Lake},
  {Quinn}, \& {Stadel}}]{ghigna_etal00}
{Ghigna}, S., {Moore}, B., {Governato}, F., {Lake}, G., {Quinn}, T., \&
  {Stadel}, J. 2000, \apj, 544, 616

\bibitem[{{Hayashi} {et~al.}(2003){Hayashi}, {Navarro}, {Taylor}, {Stadel}, \&
  {Quinn}}]{hayashi_etal03}
{Hayashi}, D., {Navarro}, J.~F., {Taylor}, J.~E., {Stadel}, J., \& {Quinn}, T.
  2003, \apj, 584, 541

\bibitem[{{Hogan} \& {Dalcanton}(2000)}]{hogan_dalcanton00}
{Hogan}, C.~J. \& {Dalcanton}, J.~J. 2000, PRD, 62, 063511

\bibitem[{{Inoue} \& {Chiba}(2003)}]{InoueChiba03}
{Inoue}, K.~T. \& {Chiba}, M. 2003, \apjl, 591, L83

\bibitem[{{Kamionkowski} \& {Liddle}(2000)}]{kamionkowski_liddle00}
{Kamionkowski}, M. \& {Liddle}, A.~R. 2000, PRL, 84, 4525

\bibitem[{{Kang} {et~al.}(2005){Kang}, {Mao}, {Gao}, \& {Jing}}]{kang_etal05}
{Kang}, X., {Mao}, S., {Gao}, L., \& {Jing}, Y.~P. 2005, Astron. \& Astrophys.
  Submitted (astro-ph/0501333)

\bibitem[{{Kauffmann} {et~al.}(1993){Kauffmann}, {White}, \&
  {Guiderdoni}}]{kauffmann_etal93}
{Kauffmann}, G., {White}, S.~D.~M., \& {Guiderdoni}, B. 1993, \mnras, 264, 201

\bibitem[{{Kawano}(2004)}]{Kawanoetal04}
{Kawano}, Y. 2004, PASJ, 56, 253

\bibitem[{{Kazantzidis} {et~al.}(2004b){Kazantzidis}, {Mayer}, {Mastropietro},
  {Diemand}, {Stadel}, \& {Moore}}]{kazantzidis_etal04b}
{Kazantzidis}, S., {Mayer}, L., {Mastropietro}, C., {Diemand}, J., {Stadel},
  J., \& {Moore}, B. 2004b, \apj, 608, 663

\bibitem[{Keeton(2001)}]{Keeton01}
Keeton, C.~R. 2001, astro-ph/0111595

\bibitem[{{Keeton}(2003)}]{Keeton03}
{Keeton}, C.~R. 2003, ApJ, 587, 143

\bibitem[{{Keeton} {et~al.}(2005){Keeton}, {Gaudi}, \&
  {Petters}}]{KeetonGaudiPetters05b}
{Keeton}, C.~R., {Gaudi}, B.~S., \& {Petters}, A.~. 2005, astro-ph/0503452

\bibitem[{{Keeton} {et~al.}(2003){Keeton}, {Gaudi}, \&
  {Petters}}]{KeetonGaudiPetters}
{Keeton}, C.~R., {Gaudi}, B.~S., \& {Petters}, A.~O. 2003, ApJ, 598, 138

\bibitem[{{Keeton} \& {Zabludoff}(2004)}]{KeetonZabludoff04}
{Keeton}, C.~R. \& {Zabludoff}, A.~I. 2004, \apj, 612, 660

\bibitem[{{Kembhavi} \& {Narlikar}(1999)}]{KembhaviNarlikar99}
{Kembhavi}, A.~K. \& {Narlikar}, J.~V. 1999, {Quasars and active galactic
  nuclei : an introduction} (Quasars and active galactic nuclei : an
  introduction /Ajit K.~Kembhavi, Jayant V.~Narlikar.~Cambridge, U.K.~:
  Cambridge University Press, c1999.~ ISBN 0521474779.)

\bibitem[{{Klypin} {et~al.}(1999){Klypin}, {Kravtsov}, {Valenzuela}, \&
  {Prada}}]{Klypinetal99}
{Klypin}, A.~A., {Kravtsov}, A.~V., {Valenzuela}, O., \& {Prada}, F. 1999,
  \apj, 522, 82

\bibitem[{{Knebe} {et~al.}(2002){Knebe}, {Devriendt}, {Mahmood}, \&
  {Silk}}]{knebe_etal02}
{Knebe}, A., {Devriendt}, J.~E.~G., {Mahmood}, A., \& {Silk}, J. 2002, MNRAS,
  329, 813

\bibitem[{{Kochanek} \& {Dalal}(2004)}]{KochanekDalal04}
{Kochanek}, C.~S. \& {Dalal}, N. 2004, \apj, 610, 69

\bibitem[{{Koopmans}(2003)}]{Koopmansetal03}
{Koopmans}, L.~V.~E. 2003, ApJ, 595, 712

\bibitem[{{Koushiappas} {et~al.}(2004){Koushiappas}, {Zentner}, \&
  {Walker}}]{koushiappas_etal04}
{Koushiappas}, S.~M., {Zentner}, A.~R., \& {Walker}, T.~P. 2004, \prd, 69,
  043501

\bibitem[{{Kravtsov} {et~al.}(2004){Kravtsov}, {Gnedin}, \&
  {Klypin}}]{kravtsov_etal04}
{Kravtsov}, A.~V., {Gnedin}, O.~Y., \& {Klypin}, A.~A. 2004, \apj, 609, 482
  (KGK04)

\bibitem[{{Libeskind} {et~al.}(2005){Libeskind}, {Frenk}, {Cole}, {Helly},
  {Jenkins}, {Navarro}, \& {Power}}]{libeskind_etal05}
{Libeskind}, N.~I., {Frenk}, C.~S., {Cole}, S., {Helly}, J.~C., {Jenkins}, A.,
  {Navarro}, J.~F., \& {Power}, C. 2005, MNRAS, submitted (astro-ph/0503400)

\bibitem[{{Lin} {et~al.}(2001){Lin}, {Huang}, {Zhang}, \&
  {Brandenberger}}]{lin_etal01}
{Lin}, W.~B., {Huang}, D.~H., {Zhang}, X., \& {Brandenberger}, R. 2001, PRL,
  86, 954

\bibitem[{{Macci\'{o}} {et~al.}(2005){Macci\'{o}}, {Moore}, {Stadel}, \&
  {Diemand}}]{Maccio_etal05}
{Macci\'{o}}, A.~V., {Moore}, B., {Stadel}, J., \& {Diemand}, J. 2005,
  astro-ph/0506125

\bibitem[{{Mao} {et~al.}(2004){Mao}, {Jing}, {Ostriker}, \&
  {Weller}}]{Maoetal04}
{Mao}, S., {Jing}, Y., {Ostriker}, J.~P., \& {Weller}, J. 2004, \apjl, 604, L5

\bibitem[{{Mao} \& {Schneider}(1998)}]{MaoSchneider98}
{Mao}, S. \& {Schneider}, P. 1998, MNRAS, 295, 587

\bibitem[{{McGaugh} {et~al.}(2003){McGaugh}, {Barker}, \& {de
  Blok}}]{mcgaugh_etal03}
{McGaugh}, S.~S., {Barker}, M.~K., \& {de Blok}, W.~J.~G. 2003, ApJ, 584, 566

\bibitem[{{Metcalf}(2004)}]{Metcalfetal04}
{Metcalf}, R.~B. 2004, ApJ, 607, 43

\bibitem[{{Metcalf}(2004a)}]{Metcalf04a}
---. 2004a, astro-ph/0407298

\bibitem[{{Metcalf}(2004b)}]{Metcalf04b}
---. 2004b, astro-ph/0412538

\bibitem[{{Metcalf} \& {Madau}(2001)}]{MetcalfMadau01}
{Metcalf}, R.~B. \& {Madau}, P. 2001, ApJ, 563, 9

\bibitem[{{Metcalf} \& {Zhao}(2002)}]{MetcalfZhao02}
{Metcalf}, R.~B. \& {Zhao}, H.-S. 2002, ApJL, 567, L5

\bibitem[{{M{\"{o}}ller}(2002)}]{Molleretal02}
{M{\"{o}}ller}, O. 2002, ApJ, 573, 562

\bibitem[{{M{\"{o}}ller} \& {Blain}(2001)}]{MollerBlain01}
{M{\"{o}}ller}, O. \& {Blain}, A.~W. 2001, MNRAS, 327, 339

\bibitem[{{M{\"{o}}ller} {et~al.}(2003){M{\"{o}}ller}, {Hewett}, \&
  {Blain}}]{MollerHewettBlain03}
{M{\"{o}}ller}, O., {Hewett}, P., \& {Blain}, A.~W. 2003, MNRAS, 245, 1

\bibitem[{{Moore} {et~al.}(1999){Moore}, {Ghigna}, {Governato}, {Lake},
  {Quinn}, {Stadel}, \& {Tozzi}}]{Mooreetal99}
{Moore}, B., {Ghigna}, S., {Governato}, F., {Lake}, G., {Quinn}, T., {Stadel},
  J., \& {Tozzi}, P. 1999, ApJL, 524, L9

\bibitem[{{Moustakas} \& {Metcalf}(2003)}]{MoustakasMetcalf03}
{Moustakas}, L.~A. \& {Metcalf}, R.~B. 2003, MNRAS, 229, 607

\bibitem[{{Mu{\~{n}}oz} {et~al.}(2001){Mu{\~{n}}oz}, {Kochanek}, \&
  {Keeton}}]{munoz_etal01}
{Mu{\~{n}}oz}, J.~A., {Kochanek}, C.~S., \& {Keeton}, C.~R. 2001, ApJ, 558, 657

\bibitem[{{Nagai} \& {Kravtsov}(2005)}]{nagai_kravtsov05}
{Nagai}, D. \& {Kravtsov}, A.~V. 2005, \apj, 618, 557

\bibitem[{{Oguri}(2004)}]{Oguri04}
{Oguri}, M. 2004, astro-ph/0411464

\bibitem[{{Quadri} {et~al.}(2003){Quadri}, {M{\"{o}}ller}, \&
  {Natarajan}}]{Quadrietal03}
{Quadri}, R., {M{\"{o}}ller}, O., \& {Natarajan}, P. 2003, ApJ, 597, 659

\bibitem[{{Rees}(1986)}]{rees82}
{Rees}, M.~J. 1986, MNRAS, 218, 25P

\bibitem[{{Refsdal}(2000)}]{Refsdaletal00}
{Refsdal}, S. 2000, A \& A, 360

\bibitem[{{Richards}(2004)}]{Richardsetal04}
{Richards}, G.~T. 2004, ApJ, 610, 679

\bibitem[{{Schechter}(2003)}]{Schechteretal03}
{Schechter}, P.~L. 2003, ApJ, 584, 657

\bibitem[{{Schechter} \& {Wambsganss}(2002)}]{SchechterWambsganss02}
{Schechter}, P.~L. \& {Wambsganss}, J. 2002, ApJ, 580, 685

\bibitem[{{Schechter} {et~al.}(2004){Schechter}, {Wambsganss}, \&
  {Lewis}}]{Schechteretal04}
{Schechter}, P.~L., {Wambsganss}, J., \& {Lewis}, G.~F. 2004, ApJ, 613, 77

\bibitem[{{Schild}(1996)}]{Schild96}
{Schild}, R.~E. 1996, ApJ, 464, 125

\bibitem[{{Schneider} {et~al.}(1992){Schneider}, {Ehlers}, \&
  {Falco}}]{Schneideretal}
{Schneider}, P., {Ehlers}, J., \& {Falco}, E.~E. 1992, {Gravitational Lenses}
  (Gravitational Lenses, XIV, 560 pp.~112 figs..~Springer-Verlag Berlin
  Heidelberg New York.~ Also Astronomy and Astrophysics Library)

\bibitem[{{Schneider} \& {Weiss}(1992)}]{SchneiderWeiss92}
{Schneider}, P. \& {Weiss}, A. 1992, A \& A, 260, 1

\bibitem[{{Shapiro} {et~al.}(1994){Shapiro}, {Giroux}, \&
  {Babul}}]{shapiro_etal94}
{Shapiro}, P.~R., {Giroux}, M.~L., \& {Babul}, A. 1994, ApJ, 427, 25

\bibitem[{{Sigurdson} \& {Kamionkowski}(2004)}]{sigurdson_kamionkowski04}
{Sigurdson}, K. \& {Kamionkowski}, M. 2004, Physical Review Letters, 92, 171302

\bibitem[{{Silk} \& {Stebbins}(1993)}]{silk_stebbins93}
{Silk}, J. \& {Stebbins}, A. 1993, Astrophys. J., 411, 439

\bibitem[{{Somerville}(2002)}]{somerville02}
{Somerville}, R.~S. 2002, ApJ, 572, 23

\bibitem[{{Somerville} \& {Primack}(1999)}]{somerville_primack99}
{Somerville}, R.~S. \& {Primack}, J.~R. 1999, MNRAS, 310, 1087

\bibitem[{{Spergel} \& {Steinhardt}(2000)}]{spergel_steinhardt00}
{Spergel}, D.~N. \& {Steinhardt}, P.~J. 2000, PRL, 84, 3760

\bibitem[{{Stoehr} {et~al.}(2003){Stoehr}, {White}, {Springel}, {Tormen}, \&
  {Yoshida}}]{stoehr_etal03}
{Stoehr}, F., {White}, S.~D.~M., {Springel}, V., {Tormen}, G., \& {Yoshida}, N.
  2003, \mnras, 345, 1313

\bibitem[{{Stoehr} {et~al.}(2002){Stoehr}, {White}, {Tormen}, \&
  {Springel}}]{stoehr_etal02}
{Stoehr}, F., {White}, S.~D.~M., {Tormen}, G., \& {Springel}, V. 2002, \mnras,
  335, L84

\bibitem[{{Tasitsiomi} \& {Olinto}(2002)}]{tasitsiomi_olinto02}
{Tasitsiomi}, A. \& {Olinto}, A.~V. 2002, PRD, 66, 083006

\bibitem[{{Taylor} \& {Babul}(2005)}]{taylor_babul05}
{Taylor}, J.~E. \& {Babul}, A. 2005, MNRAS, Submitted

\bibitem[{{Thoul} \& {Weinberg}(1996)}]{thoul_weinberg96}
{Thoul}, A.~A. \& {Weinberg}, D.~H. 1996, ApJ, 465, 608

\bibitem[{{van den Bosch} {et~al.}(2003){van den Bosch}, {Mo}, \&
  {Yang}}]{vdb_etal03}
{van den Bosch}, F.~C., {Mo}, H.~J., \& {Yang}, X. 2003, MNRAS, 345, 923

\bibitem[{{Wambsganss} {et~al.}(2004){Wambsganss}, {Bode}, \&
  {Ostriker}}]{Wambsganssetal04}
{Wambsganss}, J., {Bode}, P., \& {Ostriker}, J.~P. 2004, ApJ submitted
  (astro-ph/0405147)

\bibitem[{{White} \& {Rees}(1978)}]{white_rees78}
{White}, S.~D.~M. \& {Rees}, M.~J. 1978, MNRAS, 999, 999

\bibitem[{{Wisotzki}(2003)}]{Wisotzkietal03}
{Wisotzki}, A. 2003, A \& A, 408, 455

\bibitem[{{Wozniak}(2000)}]{Wozniaketal00}
{Wozniak}, P.~R. 2000, ApJ, 529, 88

\bibitem[{{Yoo} {et~al.}(2005){Yoo}, {Kochanek}, {Falco}, \&
  {McLeod}}]{Yooetal05}
{Yoo}, J., {Kochanek}, C.~S., {Falco}, E.~E., \& {McLeod}, B.~A. 2005, \apj,
  626, 51

\bibitem[{{Zakharov}(1995)}]{Zakharov95}
{Zakharov}, A.~F. 1995, A \& A, 293, 1

\bibitem[{{Zentner} {et~al.}(2005{\natexlab{a}}){Zentner}, {Berlind},
  {Bullock}, {Kravtsov}, \& {Wechsler}}]{zentner_etal05c}
{Zentner}, A.~R., {Berlind}, A.~A., {Bullock}, J.~S., {Kravtsov}, A.~V., \&
  {Wechsler}, R.~H. 2005{\natexlab{a}}, ApJ, 624, 505

\bibitem[{{Zentner} \& {Bullock}(2002)}]{zentner_bullock02}
{Zentner}, A.~R. \& {Bullock}, J.~S. 2002, PRD, 66, 043003

\bibitem[{{Zentner} \& {Bullock}(2003)}]{zentner_bullock03}
---. 2003, ApJ, 598, 49

\bibitem[{{Zentner} {et~al.}(2005{\natexlab{b}}){Zentner}, {Koushiappas}, \&
  {Kazantzidis}}]{zentner_etal05b}
{Zentner}, A.~R., {Koushiappas}, S.~M., \& {Kazantzidis}, S.
  2005{\natexlab{b}}, in Proceedings of the Fifth International Workshop on the
  Identification of Dark Matter

\bibitem[{{Zentner} {et~al.}(2005{\natexlab{c}}){Zentner}, {Kravtsov},
  {Gnedin}, \& {Klypin}}]{zentner_etal05a}
{Zentner}, A.~R., {Kravtsov}, A.~V., {Gnedin}, O.~Y., \& {Klypin}, A.~A.
  2005{\natexlab{c}}, ApJ, In Press, (astro-ph/0502496)

\end{thebibliography}
%

\end{document}